\documentclass[aps,prx,amsmath,amssymb,floatfix,superscriptaddress]{revtex4}
\usepackage{graphicx}
\usepackage{dcolumn}
\usepackage{bm}
\usepackage{color}
\usepackage{stmaryrd}
\usepackage{mathrsfs} 
\usepackage[dvipsnames]{xcolor}
 
\usepackage{hyperref}
\hypersetup{
     colorlinks,
    citecolor=black,
    filecolor=black,
    linkcolor=black,
    urlcolor=black,
    linktoc=all,     
    linkcolor=black,  
}

\newcommand{\be}{\begin{equation}}
\newcommand{\ee}{\end{equation}}
\newcommand{\bea}{\begin{eqnarray}}
\newcommand{\eea}{\end{eqnarray}}

\newcommand{\bK}{\mathbf{K}}

\newcommand{\rd}{\mathrm{d}}

\newcommand{\MF}{\mathrm{MF}}

\newcommand{\cC}{{\cal C}}
\newcommand{\cT}{{\cal T}}

\newcommand{\cP}{{\cal P}}

\newcommand{\cH}{{\cal H}}

\begin{document}


\title{Interacting Majorana Fermions}


\author{Armin Rahmani} 
\affiliation{Department of Physics and Astronomy \& Advanced Materials Science and Engineering Center (AMSEC),
Western Washington University, Bellingham, Washington 98225, USA}
\author{Marcel Franz}
\affiliation{Department of Physics and Astronomy \& Stewart Blusson Quantum Matter Institute,
University of British Columbia, Vancouver, British Columbia V6T 1Z1, Canada}

\begin{abstract}
Majorana fermions are the real (in a mathematical sense) counterparts of complex fermions like ordinary electrons. The promise of topological quantum computing has lead to substantial experimental progress in realizing these particles in various synthetic platforms. The realization of Majorana fermions motivates a fundamental question: What phases of matter can emerge if many Majorana fermions are allowed to interact? Here we review recent progress in this direction on the proposed experimental setups, analytical and numerical results on low-dimensional lattice models, and the exactly solvable Sachdev-Ye-Kitaev model. The early progress thus far suggests that strongly correlated phases of matter with Majorana building blocks can exhibit many novel phenomena, such as emergent spacetime supersymmetry, topological order and the physics of black-holes, in condensed matter systems. They may also provide alternative avenues for universal topological quantum computing through the realization of Fibonacci phase and measurement-based only surface codes.
\end{abstract}

\maketitle

\tableofcontents

\section{Introduction}
The emergence of several phases of electronic matter, such as high-temperature superconductors and Abelian and non-Abelian fractional quantum Hall liquids relies on strong electron-electron interactions. These phenomena cannot be understood in terms of the weakly interacting quasiparticles as in band insulators and Fermi-liquid metals. Understanding such correlated electron systems constitutes one of the most active areas of research in condensed matter physics today.

 From a theoretical perspective, electrons can be viewed as solutions to the Dirac equation, a \textit{complex} relativistic quantum mechanical equation for particles with fermionic statistics. In 1937 Ettore Majorana wrote a \textit{real} equation to describe relativistic quantum fermions \cite{Majorana1937}. Particles satisfying the Majorana equations are known as Majorana fermions. Due to the realness of the equation, these particles are their own antiparticle and have Hermitian creation and annihilation operators. Characterizing phases of matter that can emerge due to strong Majorana-Majorana interactions is a fundamental question, which until recently had remained unexplored in the condensed matter community. Indeed, unlike electrons, Majorana fermions are not intrinsic constituent of quantum materials. Thus, until recently, the question of interacting Majoranas was thought to be largely academic. 

Over the last two decades it has been realized that Majorana fermions can emerge as localized zero-energy modes in topological superconductors, exhibiting non-Abelian exchange statistics \cite{Alicea2012,Beenakker2012,Leijnse2012,Stanescu2013,Elliott2015}. A few years after the seminal proposal on topological quantum computing with non-Abelian anyons \cite{Kitaev2001}, several theoretical proposals appeared on the realization of Majorana zero modes in hybrid systems of $s$-wave superconductors deposited on the surface of a strong topological insulator \cite{FuKane2008} and  related structures  \cite{Sau2010,Alicea2010}, as well as at the endpoints of heterostructures of semiconducting nanowires  \cite{Lutchyn2010,Oreg2010}. The experimental progress following these proposals has been remarkable \cite{Mourik2012,Das2012,Deng2012,Rokhinson2012,Finck2013,Hart2014,Nadj-Perge2014,Deng2016,Zhang2018x,Jia2016a,Jia2016b,Wang2018}. The existence of unpaired localized Majoranas is now widely accepted, stimulating a worldwide effort focused on creating qubits and performing braiding with them. 

These developments open the door to studying a rich area of the physics of correlated quantum matter in a new setting. What happens when we create a large number of Majorana fermions and allow them to interact? Even in the absence of underlying strong electron-electron interactions, the emergent Majoranas can experience strong interactions relative to their kinetic energy \cite{Chiu2015}. The physics is surprisingly rich and may provide on-chip condensed matter experimental platforms to study important phenomena like spacetime supersymmetry, topological order, quantum chaos and black-hole physics. This article reviews some of the recent progress in our understanding of systems with strongly interacting Majorana fermions and highlights the remaining challenges on both the theoretical front and with respect to future experimental realizations. The field is being rapidly developed by several groups. The short list of topics discussed in this brief review was selected to give a basic introduction to the field and by no means constitutes an exhaustive review of the existing literature.

\section{Majorana vortex lattice, chiral symmetry, and strong interactions}
\subsection{Fu and Kane realization of Majorana zero modes}
The surface states of a strong topological insulator (TI) in proximity to an $s$-wave superconductor form a time-reversal-invariant analog of the $p_x+ip_y$ superconductor \cite{FuKane2008}. This in turn is a canonical platform for Majorana zero modes which are localized in the cores of Abrikosov vortices of this emergent topological superconductor.

 The simplest effective theory of the TI surface state is described by the Hamiltonian density ${\cal H}_0=\psi^\dagger(-iv\vec \sigma.\nabla-\mu)\psi$, where $\psi^\dagger=(\psi^\dagger_\uparrow,\psi^\dagger_\downarrow)$ are electron creation operators, $\vec \sigma=(\sigma_x,\sigma_y)$ are Pauli matrices, $v$ is the Fermi velocity, and $\mu$ is the chemical potential. Proximity to an $s$-wave superconductor adds an effective pairing term $\Delta_0(e ^{i\phi}\psi_\uparrow^\dagger\psi_\downarrow^\dagger+e ^{-i\phi}\psi_\downarrow\psi_\uparrow) $, where $\Delta_0$ and $\phi$ are respectively the magnitude and phase of the proximity-induced pairing.  The full Hamiltonian density can be written as
\begin{equation}\label{eq:H}
{\cal H}={1\over 2}\Psi^\dagger[-iv\tau^z\vec\sigma.\nabla-\mu\tau_z+\Delta_0(\tau^x\cos \phi+\tau^y\sin\phi)]\Psi\equiv {1\over 2}\Psi^\dagger{\mathscr H}\Psi,
\end{equation}
where $\Psi^\dagger=(\psi_\uparrow^\dagger, \psi_\downarrow^\dagger, \psi_\downarrow, -\psi_\uparrow)$ in the Nambu notation and the $\tau$ Pauli matrices mix $\psi$ and $\psi^\dagger$. Particle-hole symmetry is represented by the operator
\begin{equation}
\Xi =\sigma^y\tau^y K,
\end{equation}
where $K$ is complex conjugation. We have $\Xi^2=1$ so $\Xi^{-1}=\Xi$. It is straightforward to show that $\Xi {\mathscr H} \Xi^{-1}=-{\mathscr H} $. Therefore, the Hamiltonian satisfies particle-hole symmetry. The time-reversal operator is given by
\begin{equation}
\Theta=i\sigma^y K,
\end{equation}
with $\Theta ^2=-1$ and $\Theta^{-1}=-\Theta$. For a real superconducting order-parameter ($\phi=0$), we have $\Theta {\mathscr H} \Theta^{-1}={\mathscr H}$ and the system respects time reversal. The $\tau^y\sin \phi$ term, however, breaks time reversal as $\Theta \tau^y \Theta^{-1}=-\tau^y$.

In the absence of an external magnetic field, the superconducting order parameter can be made real ($\phi=0$). Except for the time-reversal symmetry (in the absence of external magnetic field), this system is very similar to a $p_x+ip_y$ superconductor. Applying an external magnetic field normal to the surface of the system explicitly breaks time reversal and creates Abrikosov vortices. In the presence of a vortex, e.g., $\Delta_0(r,\theta)=\Delta_0(r)$ and $\phi(r,\theta)=\theta$, the Bogoliubov de Gennes (BdG) equation for the Hamiltonian~(\ref{eq:H}) has a zero-energy solution for arbitrary $\mu$. The algebra is simple for $\mu=0$ and the quasiparticle operator corresponding to the zero-mode solution can be written as
\begin{equation}\label {eq:mode}
\gamma\propto \int d^2 r \left [e^{i\pi/4}\psi_{\downarrow}({\bf r}) +e^{-i\pi/4}\psi^\dagger_{\downarrow}({\bf r})\right] e^{-\int_0^r \Delta_0(r')dr'}.
\end{equation}

\subsection{Effective model of Majorana vortex lattice}
The zero-energy solution Eq.\ (\ref{eq:mode}) is a Majorana mode with $\gamma=\gamma^\dagger$ that is exponentially bound to the core of the vortex. Furthermore, if multiple vortices are present, we have the anticommutation relation 
\begin{equation}\label{eq:comm}
\{\gamma_i,\gamma_j\}=\delta_{ij},\quad \gamma_i=\gamma^\dagger_i,
\end{equation}
 where $i$ and $j$ label different vortices. Due to the presence of a superconducting gap, the effective low-energy theory of a vortex lattice can be formulated in terms of the above Majorana operators. Several types of terms are expected in such effective Hamiltonians: hybridization terms $it_{ij}\gamma_{i}\gamma_{j}$ originating from underlying electron hopping, and interaction-terms $g_{ijkl}\gamma_{i}\gamma_{j}\gamma_{k}\gamma_{l}$ originating from the underlying electron-electron interactions, so the effective Hamiltonian can be written as~\cite{Chiu2015}
\begin{equation}\label{eq:H_m}
H=\sum_{ij}it_{ij}\gamma_{i}\gamma_{j}+\sum_{ijkl}g_{ijkl}\gamma_{i}\gamma_{j}\gamma_{k}\gamma_{l}+\dots
\end{equation}
The hermiticity of the Hamiltonian requires both $t_{ij}$ and $g_{ijkl}$ to be real. Furthermore, we can take $g_{ijkl}$ to be fully antisymmetric as the operator $\gamma_{i}\gamma_{j}\gamma_{k}\gamma_{l}$ is fully antisymmetric and symmetric components of $g$ cancel out when summing over all Majorana sites. Although Majorana operators are Hermitian, we do not expect any terms with an odd number of Majorana modes in the effective Hamiltonian. Energy operators are generally bosonic and the underlying hopping and electron-electron interactions cannot give rise to terms with an odd number of fermionic operators. Due to the exponential factor in Eq. (\ref{eq:mode}), we expect the leading term in the Hamiltonian~(\ref{eq:H_m}) to be \textit{local}. 

\begin{figure}[h]
\includegraphics[width=5in]{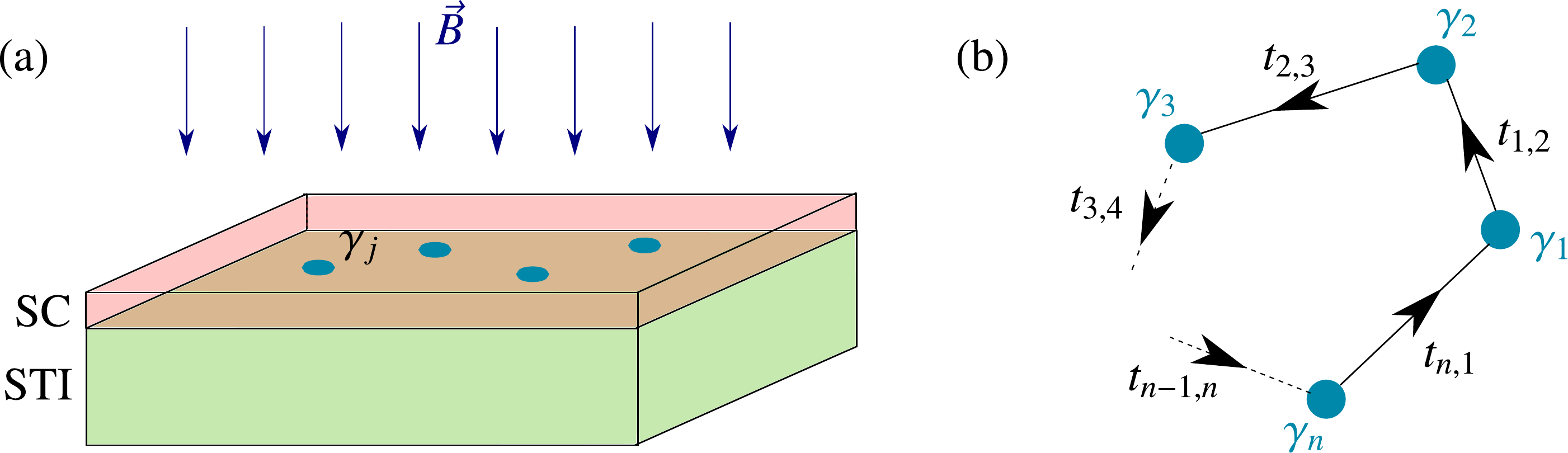}
\caption{(a) An $s$-wave superconductor in a normal magnetic field forms an Abrikosov vortex lattice. The surface state of a proximitized strong topological insulator supports Majorana zero modes exponentially bound to the vortex cores. (b) Effective low-energy description contains local hopping and four-Majorana interactions. The phase of the product of hoppings along closed polygons satisfy the Grosfeld-Stern constraint~(\ref{eq:cons}).}
\label{fig1}
\end{figure}

A more subtle constraint on the effective Hamiltonian of the vortex lattice concerns the signs of $t_{ij}$ hopping parameters. Generically, the gauge transformation
\begin{equation}
\gamma_i\to s_i\gamma_i,\ \ \  s_i=\pm 1
\end{equation}
does not change the anticommutation relations~(\ref{eq:comm}). It therefore appears that the sign of $t_{ij}$  can be gauged away. However, the product of the $t_{ij}$ around any closed path is gauge-invariant. As in Fig.~\ref{fig1}, consider a polygon whose vertices are $n$ Majorana fermions. Grosfeld and Stern~\cite{Stern2006} showed, for a of $p_x+ip_y$ superconductor, 
\begin{equation}\label{eq:cons}
\arg\left(t_{1,2}t_{2,3}\dots t_{n-1,n} t_{n,1}\right)={\pi\over 2}(n-2).
\end{equation}
The phase accumulated is due to an interplay of the order-parameter and $p$-wave pairing. Importantly, the same result applies to Majorana fermions bound to vortices in the Fu-Kane model\cite{Liu2015}.

\subsection{Chiral symmetry and strong interactions}
The effective Hamiltonian above was obtained in the regime of weak electron-electron interactions. Essentially, we worked with noninteracting surface states proximitized with quadratic mean-field pairing terms. The interaction strength $g_{ijkl}$ is therefore not expected to be large. However, due to an additional symmetry that is only present for chemical potential $\mu=0$, the Teo and Kane topological classification of defects \cite{TeoKane2010} implies that at $\mu=0$, the quadratic hopping $t_{ij}$  between the Majorana modes must vanish. Therefore by tuning the chemical potential close to zero (using gates or crystal chemistry), the hopping can be made arbitrarily small. The chemical potential does not have a strong effect on the interaction strength, which gives access to a novel strongly interacting regime.

A key idea of the Teo-Kane classification is assuming that the Hamiltonian varies slowly (away from singularities like a vortex core) so that relatively small patches in real-space have an approximate translation invariance and can be described by a momentum-dependent BdG Hamiltonian ${\cal H}({\bf k})$ labeled by the position $\bf r$ of the patch as ${\cal H}({\bf k}, {\bf r})$. For a $d$-dimensional system, $\bf k$ lives in the $d$-dimensional Brillouin zone. The positions $\bf r$ are considered on a surface that surrounds the defect. For example, in our system, $d=2$ and a loop that encloses the vortex has dimension $D=1$. The vortex in this system is a point defect with $\delta=d-D=1$.

Particle-hole and time-reversal symmetries then correspond to
\begin{equation}
{\cal H}({\bf k},{\bf r})=-\Xi {\cal H}(-{\bf k},{\bf r})\Xi^{-1}, \quad{\cal H}({\bf k},{\bf r})=\Theta {\cal H}(-{\bf k},{\bf r})\Theta^{-1}.
\end{equation}
In the presence of a vortex, the physical time reversal is broken. The system then only has a particle-hole symmetry with $\Xi^2=1$. Altland and Zirnbauer have classified the symmetry classes of noninteracting fermionic Hamiltonians with pairing~\cite{Altland1997}. In the Altland-Zirnbauer classification, the system is in the class D. Teo and Kane studied the classification of the Hamiltonian ${\cal H}({\bf k},{\bf r})$ that live on total base space $T^d\times S^D$ (the $d$-dimensional Brillouin zone and $D$ dimensional surface). By flattening the spectrum away from the energy gap between positive and negative energies, approximating $T^d\times S^D$ by $S^{d+D}$ (unimportant for strong topological invariants), and extending the notion of topological equivalence to deformation modulo the addition of trivial bands, they constructed a periodic table for various Altland-Zirnbauer symmetry classes (based on Cartan's classification of symmetric spaces) and $\delta=d-D$. Class D with $\delta=1$ has a Z$_2$ classification. This is consistent with the intuitive picture that any even number of Majorana modes can hybridize into Dirac fermions and gap out.

As discussed in Ref.\ \cite{TeoKane2010}, the neutrality point with $\mu=0$ exhibits additional symmetry and belongs to the BDI class of the Altland-Zirnbauer classification. It has a \textit{fictitious} time-reversal symmetry with $\tilde \Theta^2=1$ and consequently a chiral symmetry $\Pi=\tilde \Theta \Xi$ with $\Pi^2=1$. While the physical time-reversal is broken, the fictitious symmetry with $\tilde \Theta=\sigma^x \tau^x K$ has $\tilde \Theta {\mathscr H}\tilde \Theta^{-1}=\tilde \Theta {\mathscr H}\tilde \Theta= {\mathscr H}$ for $\mu=0$. Importantly, the BDI class with $\delta=1$ has a Z classification. Systems with $N$ vortices belong to this universality class, implying that the number of zero modes in a system is equal to the number of vortices, independently of details such as the vortex lattice geometry or spacing. This symmetry forbids hybridization between the Majorana modes, making $t_{ij}$ vanish. Small $\mu$ gives rise to small $t_{ij}\propto \mu$. Thus even with weak underlying electron-electron interactions, it is possible to tune the chemical potential of the vortex lattice close to zero and access the strongly interacting regime of large  $g/t$. This key observation, first noted in Ref. \cite{Chiu2015}, gave foundation to the recent interest in interacting Majorana models by making them potentially experimentally relevant.

\section{Majorana-Hubbard chain}
Perhaps the simplest model of interacting Majoranas is a one-dimensional chain, which can be potentially realized by a vortex lattice in a narrow strip, or by bringing together the endpoints of topological nanowires supporting Majorana fermions (reaching the strongly interacting regime may however be difficult in the latter setup). In analogy with the Hubbard model, we consider nearest-neighbor hopping and the shortest possible range interaction, which has to extend over four consecutive sites in the Majorana case. 
The Hamiltonian of the chain can be then written as~\cite{Milsted2015,Rahmani2015,Rahmani2015a}.
\begin{equation}\label{eq:RZFA}
H=\sum_j\left(it \gamma_j\gamma_{j+1}+g \gamma_j\gamma_{j+1}\gamma_{j+2}\gamma_{j+3}\right).
\end{equation}
The full phase diagram of this model has been found using a combination of field-theory and renormalization group and numerical density-matrix-renormalization-group (DMRG) calculations \cite{Rahmani2015,Rahmani2015a}. It exhibits a surprisingly rich phase diagram with several phase transitions. One of the most interesting features of the phase diagram is the realization of the tricritical Ising critical point, which exhibits emergent spacetime supersymmetry. An earlier model was constructed by Grover, Sheng and Vishwanath (GSV)~\cite{Grover2014} to realize SUSY by coupling interacting Majoranas to bosonic spins $\sigma$ sitting between Majoranas, with Hamiltonian $H=\sum_j\left[-i\gamma_j\gamma_{j+1}+ig\sigma^z_{j+1/2}\gamma_j\gamma_{j+1}+{\cal J}\sigma^z_{j-1/2}\sigma^z_{j+1/2}-h\sigma^x_{j+1/2}\right]$. The above model also realizes the tricritical Ising critical point. Importantly, upon integrating the spins, the GSV model can be viewed as a model of interacting Majoranas, albeit with long-range interactions. 
\subsection{Positive $g$}
The noninteracting point (for $g=0$) of the Hamiltonian (\ref{eq:RZFA}) is well known to be in critical phase of the transverse-field Ising model. This $1+1$-dimensional critical phase is described by a conformal field theory with central charge $c=1/2$. As two Majorana modes $\gamma_1$ and $\gamma_2$ can be combined into a Dirac fermion $c=\gamma_1+i\gamma_2$, a critical theory of free Majoranas has half the central charge of a critical theory of free Dirac fermions.

Diagonalizing the noninteracting Hamiltonian via a Fourier transformation gives a dispersion relation $E_k\propto \sin k$ with gapless point at $k=0,\pi$. The low-energy description can be constructed in terms of chiral fields with
\begin{equation}
\gamma_j(t)\approx 2\gamma_R(vt-j)+(-1)^j2\gamma_L(vt+j), \ \ \ v=4t.
\end{equation}
In terms of these low-energy degrees of freedom, the  Hamiltonian Eq.\ \ref({eq:RZFA}) is then given by
\begin{equation}
H_0=iv\int dx [\gamma_L\partial_x\gamma_L-\gamma_R\partial_x\gamma_R]-256g \int dx \gamma_R\partial_x\gamma_R\gamma_L\partial_x\gamma_L,
\end{equation}
The interaction term has RG scaling dimension 4  (1/2 for each Majorana field and 1 for each derivative) and  is  irrelevant. We therefore expect the $c=1/2$ critical phase to be stable at weak interaction for both positive and negative $g$.
We comment that there is no mass term with $m\gamma_R\gamma_L$ in the above field theory due to the translational symmetry, $\gamma_j\to \gamma_{j+1}$, which maps $\gamma_R\to \gamma_R$ and $ \gamma_L\to -\gamma_L$. Mapping the noninteracting part of the Hamiltonian to an Ising spin chain (via the Jordan-Wigner transformation) gives the self-dual point of the transverse-field Ising model. Translation in the Majorana model leads to self-duality in the Ising model, supporting criticality. In other words, the translation symmetry insures a self-duality, which will allow us to realize a tricritical point without fine tuning more than one parameter. 

\begin{figure}[h]
\includegraphics[width=5in]{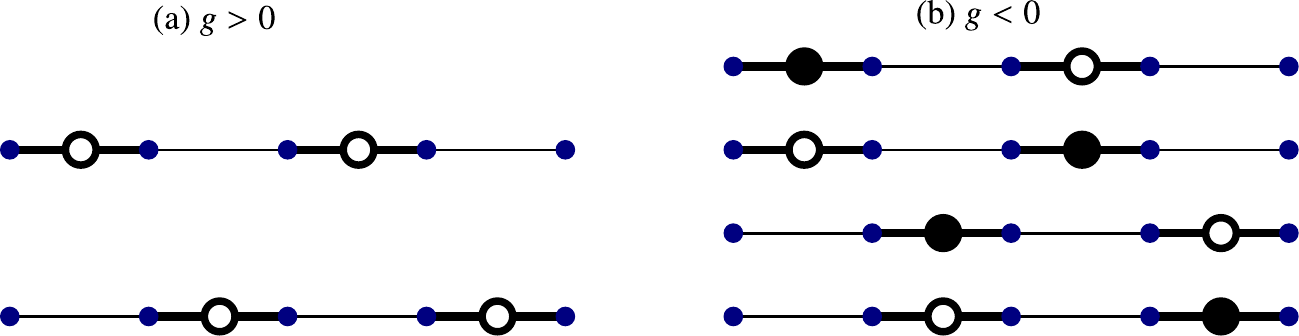}
\caption{Cartoon picture of (translation) symmetry broken phases. Small blue circles are the Majorana fermions and the larger circles on the bonds indicate Dirac fermions formed from the two Majoranas, which can be occupied or empty. (a) For $g>0$, we have shown the two states that are preferred for $t>0$. For $t<0$ we get two similar states with all Dirac fermions occupied. (b) There four degenerate states for $g<0$ remain degenerate for small $t$ regardless of its sign.}
\label{fig2}
\end{figure}

For large positive $g$, a symmetry-breaking phase transition occurs. A mean-field cartoon picture is illuminating for understanding the nature of the strongly interacting phase. Two Majorana modes can combine into a Dirac fermion $c=(\gamma_1+i\gamma_2)/2$ with $i\gamma_1\gamma_2=2c^\dagger c-1=\hat p_{12}$, where $\hat{p}=\pm 1$ is the parity of the Dirac fermion. An interaction term can be written as $g\gamma_1\gamma_2 \gamma_3\gamma_4=-gp_{12}p_{34}$. There are two ways to pair up nearest-neighbor Majoranas as shown in Fig. \ref{fig2}. For $g>0$ ($g<0$) the lower energies correspond to similar (opposite) parities. For $t=0$, we expect four-fold degenerate states for $g>0$ as the parities can be all odd (all occupied) or all even (all empty). However, an infinitesimal positive (negative) $t$ reduces the degeneracy to two-fold as it directly couples to the parities. 

These two degenerate states can be thought of as the coexistence of two phases at the self-dual point of a model with staggered interaction strength. A $c=1/2$ critical phase terminating at a gapped phase with $Z_2$ broken symmetry is a hallmark of the tricritical Ising model. It is thus expected that the transition at some finite $g$ to belong to this supersymmetric universality class.

It turns out that the actual value of $g$ at the phase transition is mysteriously large, i.e., $g/t
\sim 250$. This gives rise to a large correlation length and makes the detection of the critical point difficult through a direct numerical calculation of the central charge. However, convincing numerical evidence was obtained by studying the ratios of the spectral gaps with various boundary conditions and the exponent of correlation functions.

The finite size spectrum of the Ising and TCI model with both periodic and antiperiodic boundary conditions can be derived using the dimensions of the conformal towers and modular invariance~\cite{Cardy1986}. 
For the Ising CFT, there are three conformal towers $(\sigma ,\epsilon ,I )$,
 with the corresponding dimensions given by $(1/16,1/2,0)$. With antiperiodic boundary conditions on fermions, we have  
the operator content $\{(I,I), \quad (I,\epsilon),\quad (\epsilon, I),\quad (\epsilon, \epsilon)\}$, which  can be represented by the following expansion of the partition function in terms of the conformal towers: $Z_A=\chi_\epsilon^2+\chi_I^2+2\chi_\epsilon\chi_I$. With periodic boundary conditions, we have $Z_{P}=2\chi_\sigma^2.$

\begin{figure}[h]
\includegraphics[width=5in]{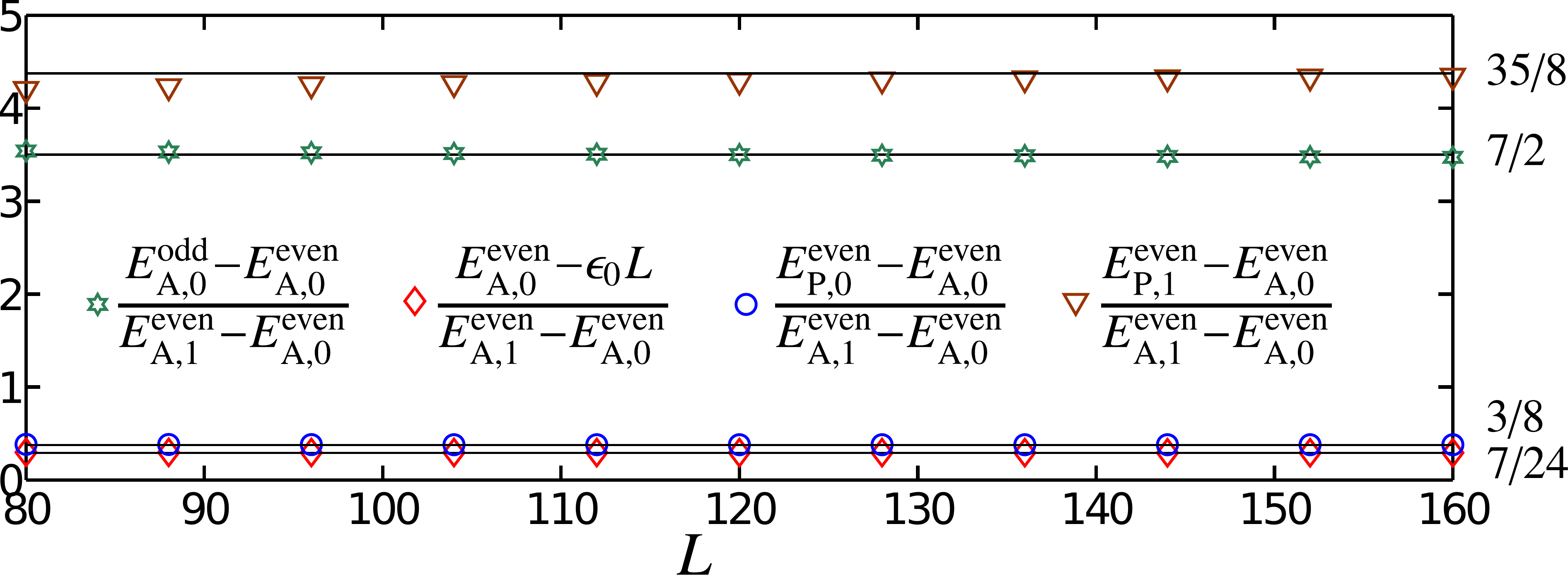}
\caption{Numerical DMRG results for universal gap ratios show perfect agreement with the CFT predictions for TCI at $t/g=0.00405$. (Figure is adapted from Ref.~\cite{Rahmani2015})}
\label{fig3}
\end{figure}

For the TCI CFT, there are six chiral conformal towers $(\epsilon ,\epsilon ',\epsilon '',\sigma ,\sigma ',I)$
 with dimensions $(1/10,3/5,3/2,3/80,7/16,0)$. We have from modular invariance ~\cite{Cappelli1987}: $Z_{A}=(\chi_\epsilon+\chi_{\epsilon '})^2+(\chi_{\epsilon '',}+\chi_I)^2$ and 
$Z_{P}=2(\chi_\sigma^2+\chi_{\sigma'}^2)$ From the above equations, we can read off the complete spectrum with periodic and antiperiodic boundary conditions for these CFTs, leading, e.g., to gap ratios:\begin{center}
\begin{tabular}{c|c|c|c|c|c}
CFT& $c$ & ~${E^{\rm odd}_{{\rm A}, 0}-E^{\rm even}_{{\rm A}, 0}\over E^{\rm even}_{{\rm A}, 1}-E^{\rm even}_{{\rm A}, 0} }$~  & ~${E^{\rm even}_{{\rm P}, 0}-E^{\rm even}_{{\rm A}, 0}\over E^{\rm even}_{{\rm A}, 1}-E^{\rm even}_{{\rm A}, 0} }$~&~${E^{\rm even}_{{\rm P}, 1}-E^{\rm even}_{{\rm A}, 0}\over E^{\rm even}_{{\rm A}, 1}-E^{\rm even}_{{\rm A}, 0} }$~&~${E^{\rm even}_{{\rm A}, 0}-\epsilon_0 L\over E^{\rm even}_{{\rm A}, 1}-E^{\rm even}_{{\rm A}, 0} }$~ \\ 
\hline 
Ising & ${1\over 2}$ & ${1\over 2}$ &  ${1\over 8}$ &  ${1\over 4}$&  ${1\over 8}$\\ 
\hline\vspace{1mm} 
~TCI~ & ${7\over 10}$ & ${7\over 2}$& ${3\over 8}$  &${35\over 8}$ & ${7\over 24}$
\end{tabular} 
\end{center}
where even and odd refer to fermion parity sectors and A and P to antiperiodic and periodic boundary conditions. Numerical calculation of these ratios provides compelling evidence for both Ising and, more importantly, the tricritical Ising phase. Some of the numerical evidence is shown in Fig.~\ref{fig3}.

\subsection{Negative $g$}
The negative $g$ side exhibits an even richer phase diagram with two phase transitions. Due to the four-fold degeneracy of the strongly interacting limit, we do not expect a TCI transition. Instead, as we increase the interaction strength, we first have a transition to an intermediate critical phase with central charge $c=3/2$. Direct numerical evidence for this change in the central charge can be obtained from calculating the entanglement entropy with DMRG.  
For a subsystem of length $y$, the entanglement entropy in a CFT with periodic boundary conditions is given by
\be S={c\over 3}\log\left[{L\over \pi a}\sin\left({\pi y\over L}\right)\right]+{\rm const.}\ee
where $L$ is the system size ~\cite{Calabrese2004}.

\begin{figure}[h]
\includegraphics[width=4in]{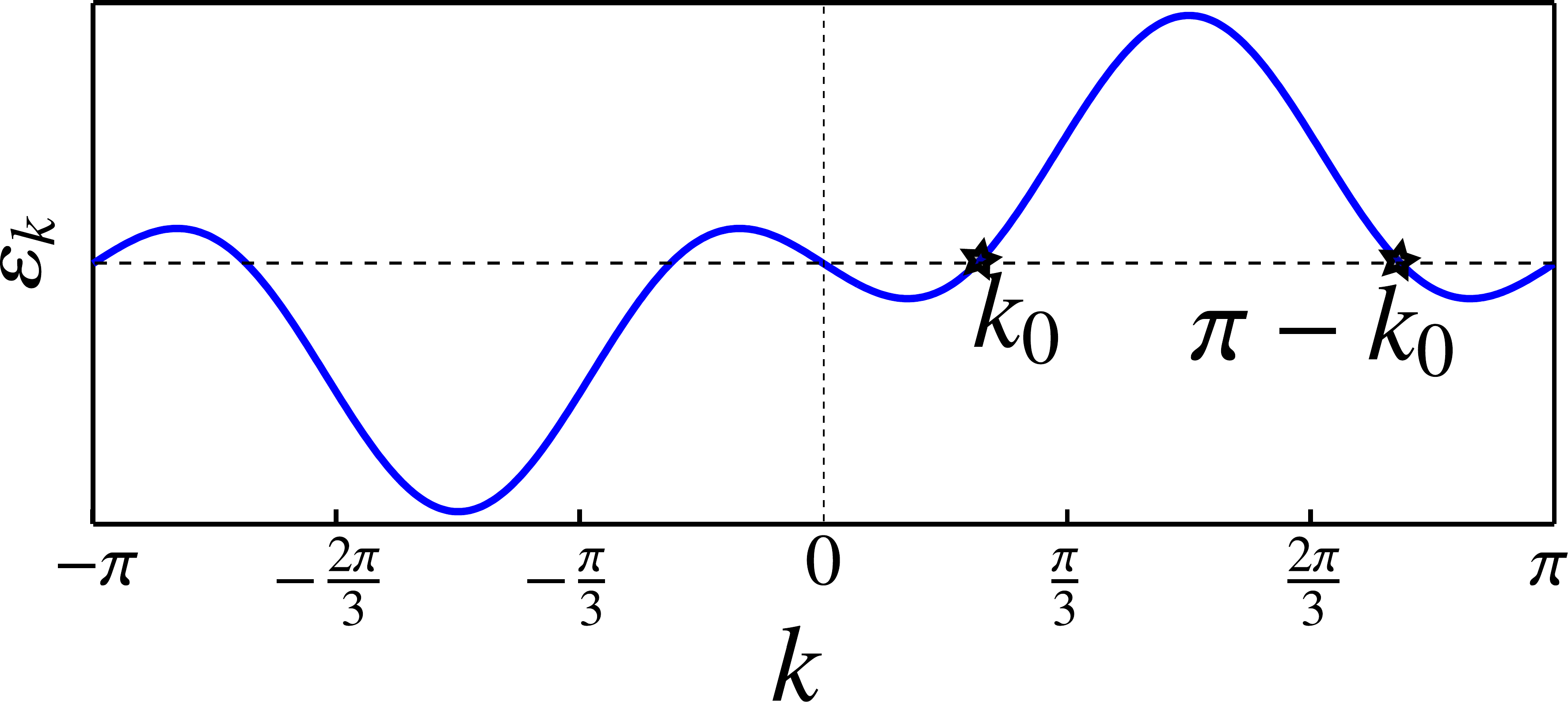}
\caption{The modified dispersion with large effective third-neighbor hopping.}
\label{fig4}
\end{figure}

It turns out that the transition between the $c=1/2$ and $c=3/2$ is a Lifshitz transition, in which the topology of the Fermi surface changes. In the one-dimensional case, the Fermi surface is a collection of nodes in the dispersion relation and a Lifshitz transition corresponds to a change in the number of nodes and therefore the number of low-energy degrees of freedom. To understand the nature of a $c=3/2$ phase, we can once again start from weakly interacting theory. We observe that third neighbor hopping is allowed by symmetry and can be generated by the interactions (second-neighbor hopping breaks the spatial parity $\gamma_j\to (-1)^{j}\gamma_{-j}$ symmetry). The presence of third-neighbor hopping can change the $t \sin k$ dispersion to a 
$\varepsilon_k=t\sin k +t'\sin 3k $ dispersion. In the presence of an effective third-neighbor hopping, the dispersion can vanish at four other points in addition to $k=0,\pi$ as shown schematically in Fig. \ref{fig4}. This introduces four new low-energy chiral Majoranas, which can be combined into two low-energy chiral Dirac fermions, i.e., one Dirac mode. The low energy expansion then reads
\be \gamma_j \approx 2\gamma_L(j)+(-1)^j2\gamma_R(j)+\left[e^{-ik_0j}\psi_R(j)+e^{i(k_0-\pi )j}\psi_L(j)+{\rm H.c.}\right].
\label{glow}\ee

The low-energy Hamiltonian of the phase in the presence of an effective third-neighbor hopping (generated due to the interactions under the RG flow) can then be written in terms of the low-energy degrees of freedom as
\be 
H=i\int dx [v_0(\gamma_L\partial_x\gamma_L-\gamma_R\partial_x\gamma_R)+v(\psi^\dagger_L\partial_x\psi_L-\psi^\dagger_R\partial_x\psi_R)]+H_{\rm int},
\ee
where the interaction terms are given by
\be H_{\rm int}= \int dx\left[g_0 :\psi^\dagger_L\psi_L\psi^\dagger_R\psi_R:+{ g'} \gamma_R\gamma_L\left(\psi_L\psi_R+\psi_L^\dagger \psi_R^\dagger \right)
\right].\ee
In the above expression ``$:$'' indicates normal ordering and $g_0=-16g[\cos k_0-\cos (3k_0)]$ for weak coupling. Therefore, negative $g$ corresponds to positive $g_0$ and repulsive interactions between the Dirac fermions. The part of the Hamiltonian that depends only on Dirac fermions is therefore described by a Luttinger liquid with Luttinger parameter $K<1$ upon bosonization:
\be
\int dx \left[iv(\psi^\dagger_L\partial_x\psi_L-\psi^\dagger_R\partial_x\psi_R)+g_0 :\psi^\dagger_L\psi_L\psi^\dagger_R\psi_R:\right]\sim \int dx \left[K(\partial_x \phi)^2+{1\over K}(\partial_x\theta)^2\right],
\ee
where $\phi$ and $\theta$ are bosonic fields with commutation relation $[\phi(x),\theta(x']=i \pi {\rm sgn}(x'-x)$. One can also demonstrate that the term which couples the Dirac fermions to the low-energy Majorana modes is \textit{irrelevant} in the RG sense. The scaling dimension of $\psi_L\psi_R$  is $1/K$ in the above Luttinger-liquid theory, and $\gamma_R\gamma_L$ has dimension 1 in the free Majorana theory, Thus the $g'$ term has a dimension $1+1/K>1$ for $K<1$ and is irrelevant. As the Luttinger-liquid sector, which has a conserved charge, is decoupled from the Majorana sector, the $c=3/2$ phase exhibits an emergent $U(1)$ charge conservation symmetry.

As shown in Fig. \ref{fig2}, in the $g\to -\infty$ limit, we expect a symmetry broken four-fold degenerate gapped phase. It appears that in the field-theory picture there are no terms, which may become relevant and cause a phase transition from the $c=3/2$ phase into a gapped phase. This apparent contradiction is resolved by considering terms that generically have fast oscillations. Such terms have been neglected in writing the low-energy interaction Hamiltonians. However, at certain commensurate momenta, the oscillations disappear and if relevant in RG, these terms can drive a phase transition.

Importantly, it appears that $k_0\to \pi /4$ and $K\to 1/4$ as we approach the phase transition and fitting to the Ising+LL begins to fail. 
This oscillatory interaction term 
\be H'\propto \int dx \gamma_R\gamma_L\left[e^{i(4k_0-\pi)x}\psi^\dagger_R\partial_x\psi^\dagger_R\psi_L\partial_x\psi_L-{\rm H.c.}\right],\ee
becomes non-oscillatory at $k_0=\pi/4$. 
The scaling dimension of $H'$ is $1+4K$. This term becomes relevant at $K=1/4$ and, as it couples both the Majorana and the Luttinger liquid sectors, is expected to gap out both sectors. The phase transition appears to be a novel generalization of the commensurate-incommensurate transition~\cite{Haldane1980,Schulz1980}. 

While the gapped phase for large negative $g$ is four-fold degenerate only in the thermodynamic limit for arbitrary boundary conditions and system size, we have an \textit{exact} degeneracy with periodic boundary condition and $L=8N$, where the number of Majoranas is $2L$. This exact degeneracy is very helpful for determining the phase transition. 

Consider the fermion parity operator 
\be
F=\gamma_0\gamma_1\ldots \gamma_{2L-1}
\ee
Translation by one site, $T$, maps $F$ to $\gamma_1\gamma_2\ldots \gamma_{2L-1}\gamma_0=-F$. As the Hamiltonian is translationally invariant, if $\psi\rangle$ is an eigenstate of energy $E$, $T|\psi\rangle$ is also an eigenstate of energy $E$. Since these two states have different fermion parity, they must be different states and each eigenstate (particularly the ground state) is at least doubly degenerate. The four-fold degeneracy of the ground state corresponds to two-fold degeneracy in a given fermion-parity sector. The argument for the exact four-fold degeneracy is more subtle and relies on an interplay of translation and spatial parity in an adiabatic interpolation between known ground states dominated by staggered interaction.

\begin{figure}[b]
\includegraphics[width=3in]{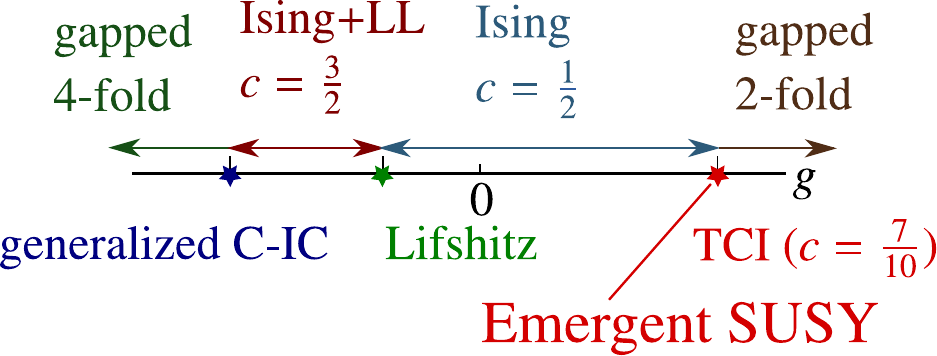}
\caption{The phase diagram of the Majorana-Hubbard chain.}
\label{fig5}
\end{figure}

The phase diagram of the model has been determined accurately and the details of the above phases are verified against DMRG calculations. In particular the effective parameters of the $c=3/2$ phase, $K$ and $k_0$, can be extracted by comparing the low-energy spectrum of the model with theoretical predictions based on the LL+Majorana picture. It was found that the commensurate-incommensurate transition (as determined from the lifting of the above-mentioned exact four-fold degeneracy)  occurs precisely at the the point where the numerically extracted $k_0=\pi/4$ and $K=1/4$. Furthermore, the Luttinger parameter $K$ extracted from the spectrum is in good agreement with the scaling behavior of the numerically calculated equal-time correlation functions in the $c=3/2$  phase, which are predicted to be
\be\label{eq:corr}
i\langle\gamma_0\gamma_{x+1}\rangle\sim a/x+b\sin\left(k_0x+\phi\right)/x^{\left(K+1/K\right)/2}.
\ee
The nature of the Lifshitz transition is also independently verified by observing that the velocities in the neighboring critical phases (the slope of the low-energy spectral gaps as a function of $1/L$) continuously approached zero. 

These analytical arguments and numerical results reveal a rich phase diagram for the interacting Hubbard-Majorana chain, the simplest possible model of interacting Majoranas, which is summarized in Fig. \ref{fig5}.

\section{Other one-dimensional models of interacting Majoranas}
In addition to the simple Majorana chain model, other one-dimensional models have been proposed~\cite{Katsura2015,Hosho,Kawabata2017}. A few one- and quasi-one-dimensional models have been proposed, which realize the same TCI supersymmetric critical point~\cite{Zhu2016,Ejima2016}. An Interesting extension of the model studied by O'Brien and Fendley~\cite{Obrien} exhibits lattice supersymmetry at interaction strengths that could be much smaller than $g/t\sim 250$ found for the Majorana-Hubbard chain. 
Here we discuss this important extension of the Majorana-Hubbard chain. Consider a Majorana chain with nearest-neighbor hopping and a four-fermion interaction described by the Hamiltonian
\be\label{eq:OF}
H=\sum_j(it\gamma_j\gamma_{j+1}+g\gamma_{j-2}\gamma_{j-1}\gamma_{j+1}\gamma_{j+2}).
\ee
Up to a constant energy shift $E_0$, we can write the above Hamiltonian as the sum of squares of two Hermitian operators
\be 
H=(Q^+)^2+(Q^-)^2-E_0,
\ee
where the Hermitian operators $Q^\pm$ are given by 
\be 
Q^\pm={1\over 2\sqrt{\lambda_3}}\sum_a(\pm1)^a(\lambda_I\gamma_a\pm i\lambda_3\gamma_{a-1}\gamma_a\gamma_{a+1}),\ee
with $\lambda_I=t/2$ and $\lambda_3=-g$. The notation follows Ref.~\cite{Obrien}, where $\lambda_I$ indicates that the quadratic term corresponds to the self-dual critical point of the transverse field Ising model (upon a Jordan Wigner transformation). Thus the quadratic terms are described by the Ising CFT.
The duality transformation of the Ising model maps to translation symmetry $\gamma_a\to\gamma_{a+1}$. It is easy to show that $Q^+$ is translationally invariant, while $Q^-\to-Q^-$ due to the $(-1)^a$ factor. Clearly, the Hamiltonian is self-dual, i.e., translationally invariant in terms of the Majoranas. 

The self-dual (translationally invariant) interactions in Eq.~\eqref{eq:OF} and the four-consecutive-site interaction of Eq.~\eqref{eq:RZFA} are both irrelevant in the renormalization group sense with respect to the Ising CFT. The least irrelevant self-dual operator in the Ising CFT is $T\bar T$, where $T$ and $\bar T$ are respectively the right- and left-moving components of the stress-energy tensor in  the CFT~\cite{Kastor}. It is reasonable to expect both these interactions to renormalize to $T\bar T$ and give the same effective field theory.

The Hamiltonian~\eqref{eq:OF} has several interesting features. First, it is \textit{exactly solvable} for $\lambda_I=\lambda_3$. As $H+E_0$ can be written as the sum of squares of two Hermitian operators, any wavefunction that simultaneously annihilates both  $Q^+$ and $Q^-$ is a ground state.  For $\lambda_I=\lambda_3=1$, we can write
\be
\begin{split}\label{eq:Qpm} 
Q^\pm\big|_{\lambda_I=\lambda_3=1}={1\over 2}\sum_j(\gamma_{2j-1}\pm \gamma_{2j+1})\left(1+i\gamma_{2j-1}\gamma_{2j}\right)
={1\over 2}\sum_j(\gamma_{2j+2}\pm \gamma_{2j-1})\left(1+i\gamma_{2j}\gamma_{2j+1}\right).
\end{split}
\ee
The ground states are easiest to understand in the Ising spin representation. A Jordan Wigner transformation maps
\be 
\gamma_{2j-1}\gamma_{2j}\to i\sigma^x_j, \quad \gamma_{2j}\gamma_{2j+1}\to i\sigma^z_j\sigma^z_{j+1}.
\ee
Notice that the quadratic term in the Hamiltonian then maps to the critical Ising chain. The two forms of $Q^\pm$ in Eq.~\eqref{eq:Qpm} indicate that three states 
\be
|G_0\rangle =|\rightarrow\rightarrow...\rightarrow\rangle, \quad |G_\uparrow\rangle =|\uparrow\uparrow...\uparrow\rangle, \quad  |G_\downarrow\rangle =|\downarrow\downarrow...\downarrow\rangle,
\ee
(all $\sigma^x$ having eigenvalue 1, and all $\sigma^z$ with eigenvalue $\pm$1) are annihilated by $Q^\pm$ and are therefore exact ground states.

As shown in Ref. \cite{Obrien}, there are no other linearly independent ground states. The $|G_\uparrow\rangle$  and $|G_\downarrow\rangle$ are ordered; they spontaneously break the $Z_2$ symmetry, while $|G_0\rangle$ is disordered. These three states coexist at a first-order phase transition if we add perturbations that break the self-duality (Majorana translation). The physics is analogous to the Majorana chain of Ref. \cite{Rahmani2015,Rahmani2015a} and a tricritical point is expected between the gapless Ising phase and the gapped phase that includes the 3-fold degenerate exactly solvable point. Detailed numerical studies based on universal gap ratios confirmed the existence of the TCI CFT at $\lambda_3\approx 0.856 \lambda_I$, which corresponds to an interaction strength of $g/t\approx 1.712$ of the order unity.

Another interesting feature of this model is that it allows for direct identification of the supersymmetric currents on the lattice. The Hamiltonian of a supersymmetric field theory in 1+1 dimensions is given by
\be 
{\cal H}=\left(\int dx  G \right)^2+\left(\int dx  \bar G \right)^2, 
\ee
where $G$ and $\bar G$ are components of the supersymmetric current. This suggests that in the scaling limit the integrals may be identified with $Q^\pm$. In Ref. \cite{Obrien}, by considering the action of duality on the currents, the following identification was made for lattice operators $G_j$ and $\bar G_j$, and the fermionic fields $\psi_j$ and $\bar \psi_j$ of the TCI CFT, and confirmed with compelling DMRG studies:
\begin{align}
G_j&=\lambda_I(\gamma_{2j-1}+\gamma_{2j})+i\lambda_3(\gamma_{2j-2}+\gamma_{2j+1})\gamma_{2j-1}\gamma_{2j},\cr
\overline{G}_j&=\lambda_I(\gamma_{2j-1}-\gamma_{2j})+i\lambda_3(\gamma_{2j+1}-\gamma_{2j-2})\gamma_{2j-1}\gamma_{2j},\cr
\psi_j&= \lambda_I(\gamma_{2j-1}-\gamma_{2j})+i\lambda_3(\gamma_{2j-2}-\gamma_{2j+1})\gamma_{2j-1}\gamma_{2j},\cr
\overline{\psi}_j&=\lambda_I(\gamma_{2j-1}+\gamma_{2j})-i\lambda_3(\gamma_{2j+1}+\gamma_{2j-2})\gamma_{2j-1}\gamma_{2j}.
\label{latticeanalogs}
\end{align}

Breaking the chiral symmetry by adding a second-neighbor Majorana hopping $-i\lambda_c\sum_j\gamma_j\gamma_{j+2}$, which changes the Hamiltonian to 
\begin{align}
\widetilde{H}&=
\left(1+\frac{\lambda_c}{\lambda_I}\right)(Q^+)^2 + \left(1-\frac{\lambda_c}{\lambda_I}\right)(Q^-)^2\cr
\label{Hchiral}
\end{align}
 leads to a rich phase diagram. The above perturbation in the Ising phase corresponds to $\int(T-\overline{T})$ in the CFT, which leaves the model conformally invariant. Numerics indicate that the gapped phase and the TCI point also remain intact for small $\lambda_c$. For $\lambda_c<\lambda_I$ the ground state in the gapped phase does not change. For $|\lambda_c|>\lambda_I$ the coefficient of one of the $(Q^\pm)^2$ terms becomes negative, leading to a phase transitions along the lines $\lambda_c=\pm\lambda_I$, along which the Hamiltonian  $2(Q^\pm)^2$ has exact lattice supersymmetry.

 \section{Fibonacci topological superconductor from interacting Majoranas}
Hu and Kane introduced a model of interacting Majoranas that gives rise to a superconducting phase described by the Fibonacci topological field theory~\cite{Hu2018}. The Fibonacci phase is of great interest as it allows for universal topological quantum computation \cite{Nayak2008_RMP,Freedman2002}. The well known proposals for achieving the Fibonacci topological phase utilize more exotic generalizations of Majorana fermions, known as parafermions~\cite{Lindner2012, Clarke2013}.  Remarkably, the Hu-Kane model realizes this phase with only interacting Majoranas.

The emergence of the TCI critical point in the two models of interacting Majoranas discussed in the previous sections suggests that interacting Majorana chain has a similar behavior to a chain of interacting Fibonacci anyons, since both a ``golden chain" model of interacting Fibonacci anyons\cite{Feiguin2007}, and the interfaces between Ising and Fibonacci topological phases \cite{Grosfeld2009} support the TCI phase.

The Hu-Kane model relies on the coset construction $SO(7)_1/(G_2)_1$, where  $SO(7)_1$ describes $7$ free chiral Majorana fermions with central charge $c=7/2$ and $(G_2)_1$ describes the Fibonacci CFT with $c=14/5$. The central charge of the quotient above is $c=7/2 - 14/5 = 7/10$  describing the TCI CFT. This construction allows one to decompose 7 chiral Majorana modes into a Fibonacci and a TCI CFT. It is well known that the exceptional Lie algebra $G_2$ supports Fibonacci anyons at level one. Furthermore, $G_2$ is a subgroup of $SO(7)$, which preserves the octonion algebra. Octonions are generalizations of complex numbers (described by two real numbers) to objects  $q = q_0  + \sum_{a=1}^7 q_a e_a$ described by eight real numbers with $\bar q q=q_0^2+\sum_{a=1}^7 q^2_a$. The preservation of the norm is guaranteed by
\begin{equation}
e_a e_b = - \delta_{ab} + C_{abc} e_c,  
\label{octonion multiplication}
\end{equation}
where $C_{abc}$ is a totally antisymmetric tensor. The seven unit vectors $e_a$ transform under $SO(7)$ rotation. However, only a subset of rotations, i.e., $G_2$, preserve $C_{abc}$. While $SO(7)$ has 21 generators, $G_2$ has 14 generators, which can be represented by $7\times 7 $ matrices $M^A$, $A=1\dots 14$, with
\begin{equation}
\sum_A M^A_{ab} M^A_{cd} = \frac{2}{3}(\delta_{ad}\delta_{bc}-\delta_{ac}\delta_{bd}) - \frac{1}{3} *C_{abcd}.
\label{casimir}
\end{equation}
Here $*C_{abcd} = \epsilon_{abcdefg}C_{efg}/6$ is the dual of $C_{abc}$.

The coset construction starts from 7 chiral free Majoranas with
\begin{equation}
H_0 = -\frac{i v}{2}\sum_{a=1}^7 \gamma_a \partial_x \gamma_a.
\end{equation}
The $(G_2)_1$ currents can be written as 
\begin{equation}
J^A = \sum_{ab} \frac{1}{2}M^A_{ab} \gamma_a \gamma_b, \quad A=1\dots 14.
\label{J^A}
\end{equation}
and the  Sugawara form of the Fibonacci sector Hamiltonian can be written as 
 \begin{equation}
H_{\rm FIB} = \sum_A \frac{\pi v }{5}J^A J^A.
\label{J^A}
\end{equation}
Using (\ref{casimir}), one can then find the decoupling $H_0=H_{\rm FIB}+H_{\rm TCI} $ with
\begin{align}
&H_{\rm FIB}  = -\frac{2 i v}{5}\sum_a\gamma_a\partial_x\gamma_a - \frac{\pi v}{60}\sum_{abcd} *C_{abcd}\gamma_a\gamma_b\gamma_c\gamma_d, \nonumber\\
 &H_{\rm TCI} =  -\frac{i v}{10}\sum_a \gamma_a\partial_x\gamma_a +\frac{\pi v}{60} \sum_{abcd} *C_{abcd}\gamma_a\gamma_b\gamma_c\gamma_d.
\end{align}
The Hamiltonian $H_{\rm FIB}$ describes a $(G_2)_1$ CFT, with two primary fields $1$, $\tau$ of dimension $h= 0$, $2/5$.

\begin{figure}[h]
\includegraphics[width=5in]{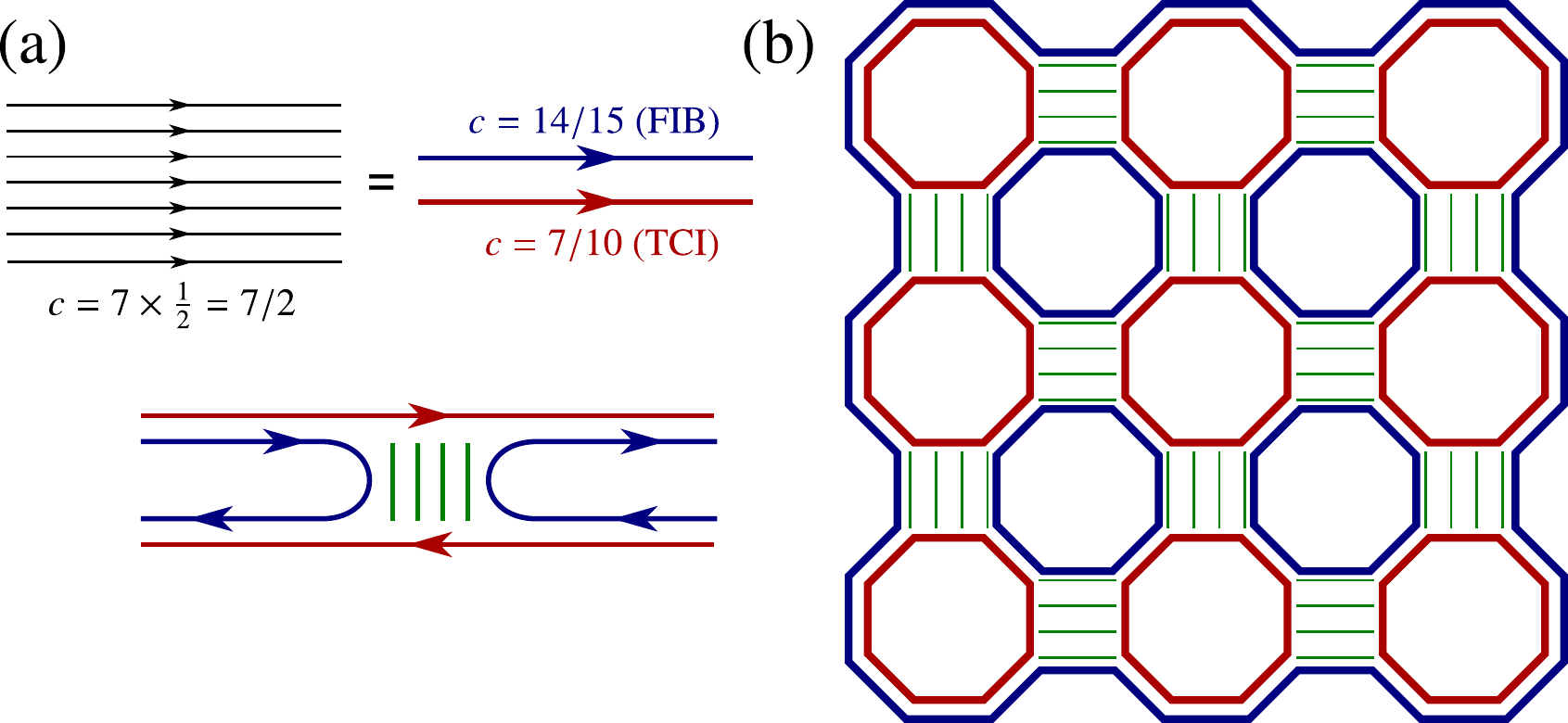}
\caption{(a) The coset construction for decoupling a $c=7/2$ CFT of 7 chiral Majoranas into a $G(2)_1$ Fibonacci CFT with $c=14/15$ and a TCI CFT with $c=7/10$. An interaction that couples right- and left-moving Fibonacci currents (shown with green lines) gaps out the Fibonacci sector. Incoming Majoranas incident on a region with this interaction then split, with the TCI field $\epsilon$ transmitted and the FIB field $\tau$ reflected. (b) A network construction with an arrangement of superconducting islands and regions with the Fibonacci current interactions, which yields a gapped phase with a Fibonacci edge mode.   Figures are based on Fig. 1 of Ref. \cite{Hu2018}. See Ref. \cite{Hu2018} also for a similar construction of a novel anti-Fibonacci phase.}
\label{fig:fib}
\end{figure}

The Majorana fermion $\gamma_a$ factors into the product 
\begin{equation}
\gamma_a = \tau_a \times \epsilon,
\end{equation}
where $\epsilon$ is a primary operator of the TCI with dimension $h=1/10$. Coupling the right-moving and left moving Fibonacci currents gaps out the Fibonacci sector, while the TCI sector remains gapless. Thus, introducing this interaction in a region using the Hamiltonian
\begin{equation}
H = -\frac{i v}{2}\sum_a (\gamma_{aR}\partial_x\gamma_{aR} - \gamma_{aL}\partial_x\gamma_{aL}) + \lambda\sum_A J^A_R J^A_L,
\label{hlambda}
\end{equation}
causes 7 incident chiral Majorana modes to factorize into an $\epsilon$ and a $\tau$, with $\tau$ reflected and $\epsilon$ transmitted, as shown in Fig. \ref{fig:fib}(a). Hu and Kane presented a two-dimensional network of superconducting islands supporting Majorana edge modes, with the Fibonacci current-current interaction present in certain regions between the islands. As illustrated in Fig. \ref{fig:fib}(b) the result is a gapped Fibonacci phase with a surviving $\tau$ edge mode.

\section{Majorana-Hubbard Ladders}

\begin{figure}[]
	\includegraphics[width=.95\columnwidth]{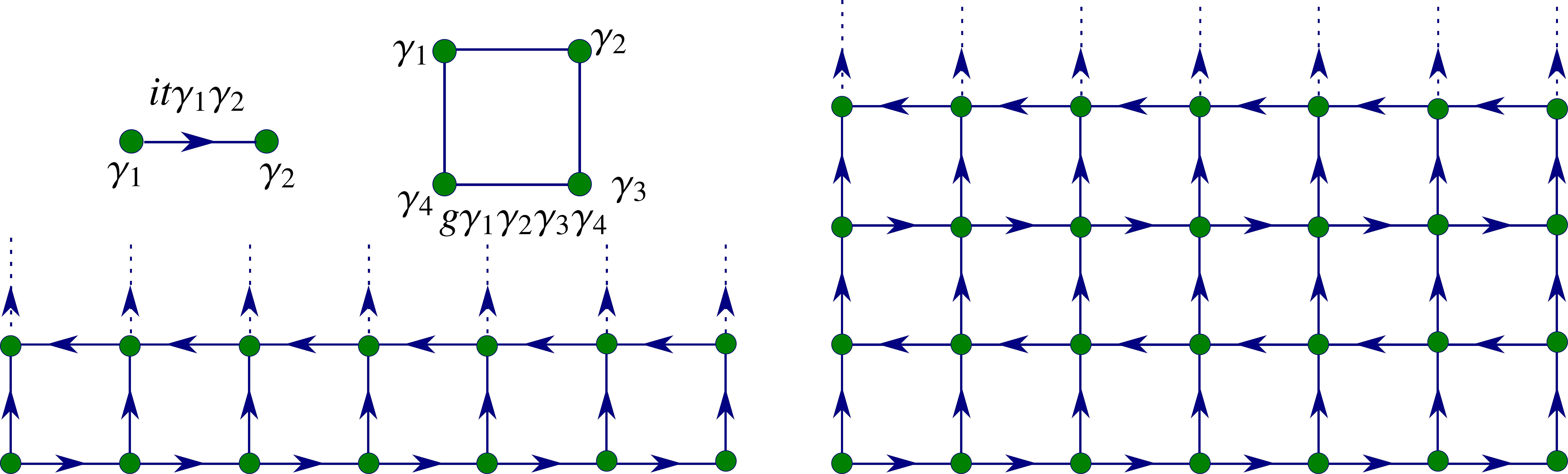}
	\caption{A schematic of the two-leg and four-leg ladders with periodic boundary conditions in the $y$ direction. The arrows indicate a Majorana hopping $it\gamma_1\gamma_2$ and there is a four-Majorana interaction term in each square plaquette.\label{fig:ladder} }
\end{figure}
An important extension of the Majorana-Hubbard chain is to a Majorana-Hubbard ladder, which includes several coupled chains of Majoranas. The simplest ladder has only two legs. Recently, significant progress was made in understanding the phases of the two- and the four-leg ladders~\cite{Rahmani2018}. Interestingly, the two-leg ladder maps onto a well understood spin chain and exhibits much simpler physics than the Majorana-Hubbard chain. Increasing the number of legs leads to richer physics. Particularly, the four-leg ladder has a novel phase diagram with multiple phases and phase transitions. 
The ladder model with shortest distance interaction has four-Majorana interactions on plaquettes, as shown in Fig. \ref{fig:ladder}. The signs of nearest-neighbor hopping between Majorana modes have signs satisfy the Grosfeld-Stern constraint~\cite{Grosfeld2006}, namely, the product of the hopping amplitudes when going around a plaquette is equal to -1. The Hamiltonian for $m$ legs can be written as
\be \label{eq:H0}H=it\sum_m\sum_{n=1}^W\gamma_{m,n}[(-1)^n\gamma_{m+1,n}+\gamma_{m,n+1}]+g\sum_{m,n}\gamma_{m,n}\gamma_{m+1,n}\gamma_{m+1,n+1}\gamma_{m,n+1},\ee
where we assume periodic boundary conditions in the $y$ direction, $\gamma_{m,n+W}=\gamma_{m,n}$. The dashed lines above the ladder in Fig. \ref{fig:ladder} indicate vertical hoppings as well as plaquette interactions between the top and the bottom rows of Majoranas.

\subsection{Two-leg ladder}
We start by writing the two-leg model in terms of Dirac fermions
\be c_{m}\equiv i^m{\gamma_{m,0}+i(-1)^m\gamma_{m,1}\over 2}.\label{c2leg}\ee
Noting that in the two-leg case, the vertical hoppings cancel out due to the periodic boundary condition in the $y$ direction, we can write the Hamiltonian as
\be H=-2t\sum_m[c^\dagger_mc_{m+1}+c^\dagger_{m+1}c_m]+2g\sum_m(2c^\dagger_mc_m-1)(2c^\dagger_{m+1}c_{m+1}-1),\ee
in terms of the Dirac fermions~\eqref{c2leg}. Upon a Jordan-Wigner transformation, this model maps to the well known XXZ spin chain:
\be H=\sum_m[t(\sigma^1_m\sigma^1_{m+1}+\sigma^2_m\sigma^2_{m+1})+2g\sigma^3_m\sigma^3_{m+1}],\ee

For $|g|<1/2$ a Luttinger-liquid phase occurs. At $g=\pm 1/2$, we have transitions to ordered phases. Since
\be i\gamma_{m,0}\gamma_{m,1}=(-1)^m(2c^\dagger_mc_m-1)=(-1)^m\sigma^3_m,\label{FAF}\ee
the  ordered phases are ferromagnetic or antiferromagnetic order in terms of $\sigma^3$, which is in turn related to the occupation number of the Dirac fermions $c_m$ formed by Majorana modes on vertical bonds through $\sigma^3_m=2c_m^\dagger c_m-1$.

\subsection{Four-leg ladder}

We now turn to the more complicated four-leg ladder. We start by considering the weak coupling limit. Due to the alternating sign of the horizontal hopping, it is convenient to introduce $e/o$ labels for the Majorana fermions on even and odd rows:
\be \gamma_{2j}\equiv \gamma^e_{2j},\ \  \gamma_{2j+1}\equiv \gamma^o_{2j+1}.\ee
For a chain of $2W$ rows of length $L$ we ca Fourier transform the Majorana operators as
\be \gamma^e_{\vec k}\equiv {1\over \sqrt{2WL}}\sum_{m,n}e^{i(mk_x+2nk_y)}\gamma^e_{m,2n},\ \ 
\gamma^o_{\vec k}\equiv {1\over \sqrt{2WL}}\sum_{m,n}e^{i(mk_x+(2n+1)k_y)}\gamma^0_{m,2n+1}.
\ee
The hopping term in $H$ can then be written as
\be H_0=-4t\sum_{k_x>0,k_y}\left[\left(\gamma^{e\dagger}_{\vec k}\gamma^e_{\vec k}-\gamma^{o\dagger}_{\vec k}\gamma^o_{\vec k}\right)\sin k_x
+\left(\gamma^{e\dagger}_{\vec k}\gamma^o_{\vec k}+\gamma^{o\dagger}_{\vec k}\gamma^e_{\vec k}\right)\sin k_y\right] .\label{H0}\ee
Using $\gamma^{e/o}_{-\vec k}=\gamma^{e/o\dagger }_{\vec k}$, we can restrict the Brillouin zone to $0\leq k_x<\pi$, $-\pi /2\leq k_y<\pi /2$. 
The energy bands are:
\be E_{\pm}=\pm 4t\sqrt{\sin^2 k_x+\sin^2k_y}.\label{E0}\ee
For the four-leg ladder, we have $W=2$ and $L\to\infty$.
With $W=2$, there are only two allowed values of $k_y=0,\pi/2$, and only $k_y=0$ is gapless.
Keeping this mode only, gives a low-energy theory identical to the two-leg case, since the two-leg ladder in this description has one allowed momentum $k_y=0$. Thus, in the low-energy limit and no interactions, the four-leg ladder reduces to the two-leg ladder. Thus, for weak interactions, we obtain the same massless Luttinger liquid phase as the two-leg ladder.

We now consider the strong-coupling limit of the four-leg ladder.
The interaction Hamiltonian can be written as
\be H_{int}=-g\sum_m\sum_{j=0}^3(i\gamma_{m,j}\gamma_{m,j+1})(i\gamma_{m+1,j}\gamma_{m+1,j+1})\ee
Without hopping, the fermion parity on each rung is conserved. The four Majoranas on each rung can be mapped to two Dirac fermions
\bea c_{m,1}&\equiv& (\gamma_{m,0}+i\gamma_{m,1})/2\nonumber \\
c_{m,2}&\equiv& (\gamma_{m,2}+i\gamma_{m,3})/2.\eea
This gives a four-dimensional Hilbert space for each rung. As the fermion parity is conserved for each rung, we can consider the two parity sectors separately and model each rung as a two-level system:
\bea
{\rm Even~parity}:\quad |\downarrow \rangle \equiv |0\rangle,\quad 
|\uparrow \rangle \equiv c_1^\dagger c_2^\dagger |0\rangle,\\
{\rm Odd~parity}:\quad |\downarrow \rangle \equiv c_1^\dagger|0\rangle,\quad 
|\uparrow \rangle \equiv  c_2^\dagger |0\rangle.
\eea
Furthermore, the Majorana bilinears can be identified with 
\bea
{\rm Even~parity}:i\gamma_{m,0}\gamma_{m,1}=  i\gamma_{m,2}\gamma_{m,3}=\sigma^3_m,\quad 
i\gamma_{m,1}\gamma_{m,2}=-i\gamma_{m,3}\gamma_{m,0}=\sigma^1_m\\
{\rm Odd~parity}: -i\gamma_{m,0}\gamma_{m,1}=  i\gamma_{m,2}\gamma_{m,3}=\sigma^3_m,\quad 
i\gamma_{m,1}\gamma_{m,2}=i\gamma_{m,3}\gamma_{m,0}=-\sigma^1_m
\eea
Assuming the same fermion parity on every rung, the interaction Hamiltonian maps to the following spin chain: 
\be H_{\rm int}=-2g\sum_m(\sigma^3_m\sigma^3_{m+1}+\sigma^1_m\sigma^1_{m+1})\ee
The above model maps to gapless free fermions. Unlike the two-leg case, the four-leg model is massless at $t=0$. 

Adding a small hopping term to the $t=0$ model, we find an effective Hamiltonian
 \bea H&=&\sum_m[-2g(\sigma^3_m\sigma^3_{m+1}+\sigma^1_m\sigma^1_{m+1})+2t\sigma^3_m]\ \  (\hbox{even parity)}\nonumber \\
 &=&\sum_m[-2g(\sigma^3_m\sigma^3_{m+1}+\sigma^1_m\sigma^1_{m+1})-2t\sigma^1_m]\ \ (\hbox{odd parity)}.\eea

For $g>0$,  we find ferromagnetic states with explicitly broken symmetry with $\langle \sigma^3_m\rangle <0$ for even fermion parity and $\langle \sigma^1_m\rangle >0$ for odd fermion parity. 
For $g<0$, we expect an in-plane antiferromagnetic order with $\langle \sigma^1_m\rangle \propto (-1)^m $ ( $\langle \sigma^3_m\rangle \propto (-1)^m$) for even (odd) fermion parity.

\begin{figure}[]
	\includegraphics[width=.9\columnwidth]{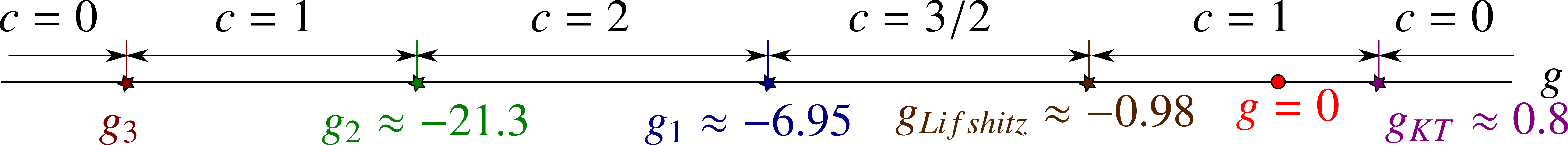}
	\caption{The phase diagram of the four-leg ladder with periodic boundary conditions in the $y$ direction. The figure is from Ref.~\cite{Rahmani2018}.  \label{fig:ladder_PD}}
\end{figure}

The full phase diagram of the model, consistent with the weak and strong-coupling descriptions, has been obtained numerically~\cite{Rahmani2018}. A direct calculation of the central charge (using the entanglement entropy) reveals the structure of the phase diagram. The results are shown in Fig.~\ref{fig:ladder_PD}.
The gapped phases with $c=0$ at $g\to\pm \infty$ are the ferromagnetic and antiferromagnetic ordered states described above for strong coupling. The $c=1$ phase that extends to both sides of the noninteracting point $g=0$ is the Luttinger-liquid phase predicted above. On the positive $g$ side only one transition occurs before the strong coupling gapped phase. On the the negative $g$ side, on the other hand, three transition occurs, with the central charge going through $3/2$, $2$, and $1$ before dropping to $c=0$ at strong coupling. 

The nature of the two phase transitions around the Luttinger-liquid phase are well understood. The transition on the positive $g$ side is of Kosterlitz-Thouless type. For small $g/t$, the $k=0$ sector is in a gapless Luttinger liquid, while the $k=\pi$ sector is gapped. Replacing the $k=\pi$ operators by their average, the following $U(1)$ symmetry-breaking perturbation was derived in Ref.\ \cite{Rahmani2018}:
 \be \delta H'=\sum_{m,m'}(2g_{m-m'}-\lambda_{m-m'})(\sigma^+_{m,0}\sigma^+_{m+1,0}\sigma^+_{m',0}\sigma^+_{m'+1,0}+h.c.).\ee
Upon bosonization, the  most relevant term comes from the
staggered part of the spin operators $\sigma^+_{m,0}\propto (-1)^me^{i\sqrt{\pi /K}\theta}$.
 Thus we get a bosonized symmetry breaking interaction:
 \be \delta H'\propto \int dx \cos [4\sqrt{\pi /K}\theta (x)]\ee
 of scaling dimension $d=4/K$. 
 For $g > 0$, this interaction becomes more irrelevant by increasing $g$, while for $g < 0$ it becomes more relevant. It appears that on the positive $g$ side, we do not have any terms that can drive a phase transition. However, the standard  Umklapp term $\psi^\dagger_L\partial_x\psi^\dagger_L\psi_R\partial_x\psi_R+h.c. \propto \cos (4\sqrt{\pi K}\phi )$, which becomes relevant at $K=1/2$ an cause a Kosterlitz-Thouless phase transition.
Numerical calculation of the Majorana two-point functions with DMRG support this picture~\cite{Rahmani2018}. On the negative $g$ side, although we have an interaction, which becomes more relevant upon increasing the interaction strength, the first phase transition occurs by a different mechanism. Similar to the Majorana-Hubbard chain, the velocity of the critical $c=1$ phase is continuously decreasing when increasing the negative interaction strength. This velocity reaches zero, as directly observed by DMRG numerics~\cite{Rahmani2018}, and drives a a Lifshitz transition at $g\approx -0.98$.

\section{Majorana-Hubbard model on the square lattice}
Two-dimensional models of interacting Majoranas have also been  studied. Although the powerful bosonization and DMRG methods cannot be applied to two-dimensional systems in the thermodynamic limit, some progress has been made by studying the weak- and strong-coupling limits as well as various mean-field models. In some limits, quantum Monte Carlo has also been applied to the two-dimensional models of interacting Majoranas~\cite{Hayata2017,Kamiya}.
The simplest two-dimensional model of interacting Majoranas is the square lattice model~\cite{Affleck2017,Kamiya}. We can assume nearest-neighbor hopping between Majorana modes and choose the signs such that the Grosfeld-Stern constraint is satisfied. The shortest distance four-fermion interaction on the square lattice corresponds to the product of the four Majorana modes along a plaquette. Therefore, we can write the Hamiltonian
\be \label{eq:H0}H=it\sum_{m,n}\gamma_{m,n}[(-1)^n\gamma_{m+1,n}+\gamma_{m,n+1}]+H_{\rm int},\ee
where $\gamma_{m,n}$ is at the lattice point  $m\hat x+n\hat y$ and 
\be H_{\rm int}=g\sum_{m,n}\gamma_{m,n}\gamma_{m+1,n}\gamma_{m+1,n+1}\gamma_{m,n+1}.\ee
Again, the realizations of this model can occur in a vortex lattice, in which the microscopic details favor a square-lattice arrangement of Abrikosov vortices~\cite{Chiu2015}.  Short-range Majorana-Majorana interactions may also occur in He$^3$~\cite{Park2015}
It is sufficient to consider only $t>0$ as the sign of $t$ can change by a $Z_2$ gauge transformation. We will consider both signs of $g$. 

The model has been studied both from the weak coupling (using a combination of field theory, renormalization group 
and mean-field approaches)~\cite{Affleck2017} and the strong coupling~\cite{Kamiya} sides. While a direct numerical determination of the phase diagram like the one-dimensional case is not possible for the full phase diagram of this model, the two complementary studies reveal many important properties of this model.

\subsection{Strong coupling}

\begin{figure}[]
\includegraphics[width=4in]{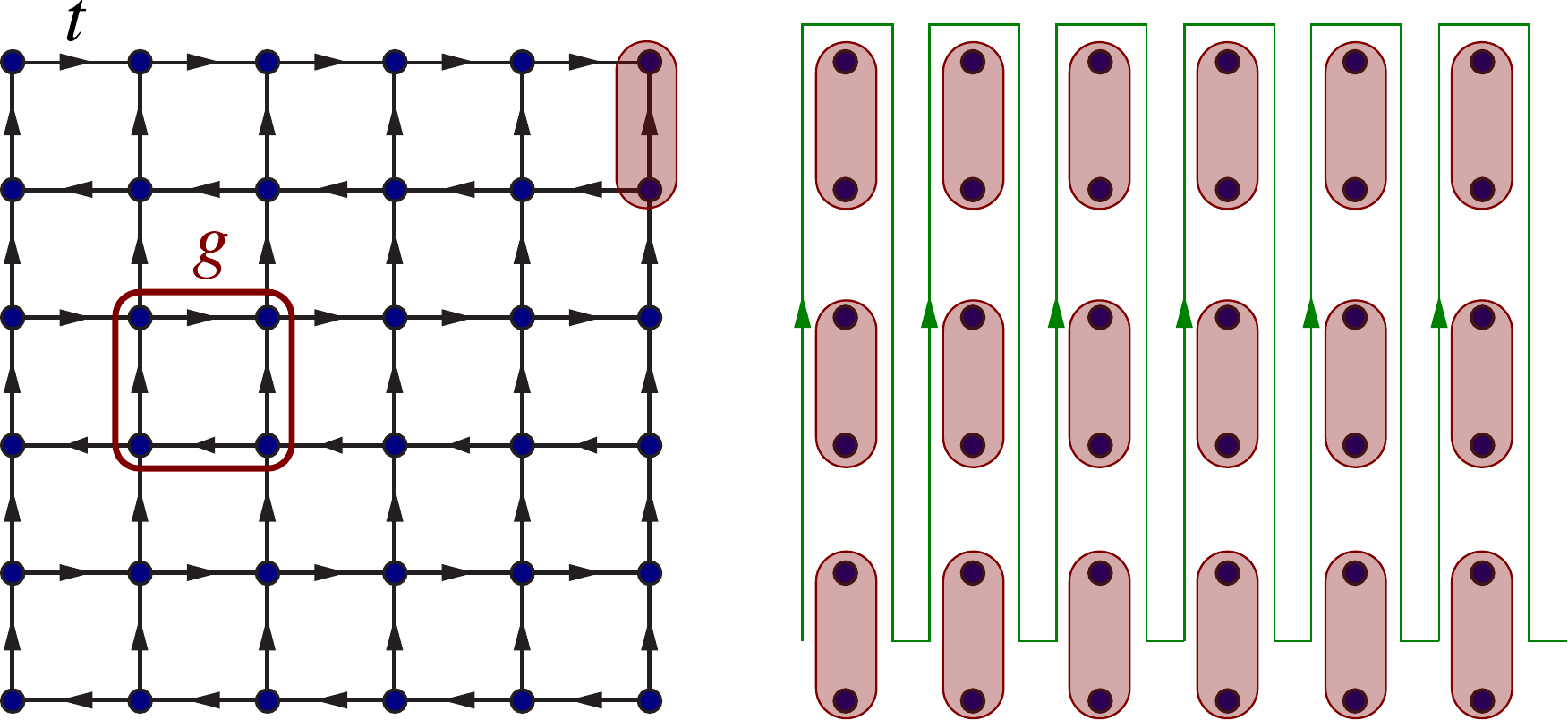}
\caption{The Majorana-Hubbard model on the square lattice and with nearest-neighbor hopping and four-Majorana interactions on plaquettes. The mapping to a spin model relies on pairing up Majoranas into Dirac fermions (pink ellipses), and using a Jordan-Wigner string that snakes through the system (green directed path). For the plaquette interaction, the string operators cancel out and we obtain a local spin model. (Figure adapted from Ref.~\cite{Kamiya})}
\label{fig6}
\end{figure}

This model, in general, suffers from the sign problem (unlike several models with two species of Majorana modes, which have a conserved particle number and were recently found to be amenable to sign-problem-free quantum Monte Carlo~\cite{Xiang2015,Hayata2017}).
Interestingly, the model is sign-problem-free for the hopping amplitude $t=0$~\cite{Kamiya}. Setting $t=0$, we first pair up two nearest-neighbor Majoranas connected by a vertical bond into  Dirac fermions and then use a Jordan-Wigner transformation along a one-dimensional path that snakes through the system as shown in Fig. \ref{fig6}. The procedure gives the following Hamiltonian for $t=0$.
\begin{equation}\label{eq:hjp}
H= -J \sum_{m,n} \hat{\sigma}_{m,n}^z \hat{\sigma}_{m+1,n}^z
  - P \sum_{m,n} \hat{\sigma}_{m,n}^x  \hat{\sigma}_{m+1,n}^x \hat{\sigma}_{m+1,n+1}^x \hat{\sigma}_{m,n+1}^x,  \quad P=J=g,
\end{equation}
which includes $\sigma^z\sigma^z$ coupling in the horizontal directions as well as the products of four $\sigma^x$ around plaquettes. We note that due to the pairing of Majorana models into Dirac modes, the spin live on and $X\times (Y/2)$ dimensional lattice for and $X\times Y$ Majorana lattice.

Remarkably, the above spin Hamiltonian for $t=0$ is sign-problem free. Monte Carlo studies in Ref. \cite{Kamiya} show a finite-temperature second-order phase transition. To understand the nature of the low-temperature phase, it is illuminating to consider two other transformations, which reveals a mapping to two decoupled copies of the quantum compass model, each living on an $(X/2)\times (Y/2)$ dimensional lattice. First, we can define spins $\tau$ living on horizontal bonds between $\sigma$ spins and then spins $\mu$ living on vertical bonds between the $\tau$ spins
\be
\tau^z_{m,n}=\sigma^z_{m,n}\sigma^z_{m+1,n},\quad
\mu^x_{m,n}=\tau^x_{m,n}\tau^x_{m,n+1} \ee
It is clear that the plaquette operator in Hamiltonian (\ref{eq:hjp}), which flips four $\sigma^z$ spins corresponds to $\tau^x_{m-1,n}\tau^x_{m+1,n}\tau^x_{m-1,n+1}\tau^x_{m+1,n+1}$. Thus the even and odd columns in the lattice of the $\tau$ decouple. We can then write the Hamiltonian as $H+H_e+H_o$, where each of the decoupled Hamiltonians is defined on a square lattice of the $\mu$ spins with $\mu^x\mu^x$ ($\mu^z\mu^z$) coupling in the $x$ ($y$) direction:
\be
H_{e/e}=\sum_{ij}(-P\mu^x{i,j}\mu^x{i+1,j}-J\mu^z_{i,j}\mu^z{i,j+1})
\ee
The quantum compass model has a long history and exhibits nematic phase with Majorana stripes~\cite{Mishra2004,Tanaka2007,Wenzel2008}, which was also confirmed by the Monte Carlo simulations of Ref.~\cite{Kamiya}.

\subsection{Mean-field theory}
With the hopping parameter, the stripe physics is simply captured by a mean-filed study. Interestingly, the stripe phases have many similarities with the symmetry breaking phases of the one-dimensional model. The interaction term in the original Majorana-Hubbard model can be factorized into horizontal, vertical, or diagonal nearest-neighbor factors:
\bea H_{int}&=&g\sum_{m,n}(i\gamma_{m,n}\gamma_{m+1,n})(i\gamma_{m,n+1}\gamma_{m+1,n+1})
=-g\sum_{m,n}(i\gamma_{m,n}\gamma_{m,n+1})(i\gamma_{m+1,n}\gamma_{m+1,n+1})\nonumber \\
&=&-g\sum_{m,n}(i\gamma_{m,n}\gamma_{m+1,n+1})(i\gamma_{m,n+1}\gamma_{m+1,n})\label{Hintfac}
\eea
The horizontal and vertical order parameters are related by a $\pi /2$ rotation and we do not need to consider them separately. It turns out that in the strong coupling regime, the diagonal decoupling yields higher energies and will not be considered further in this review (although in intermediate regimes, it may provide the lowest-energy mean-filed state as discussed in Ref. \cite{Affleck2017}).

Focusing on the vertical mean-field decoupling, we can write the expectation value $O_{m,n}\equiv \langle i\gamma_{m,n}\gamma_{m,n+1}\rangle$,
We also make the additional assumption that assume that 
\be \label{eq:O}
O_{m,n}=A+B(-1)^n+C(-1)^m+D(-1)^{m+n}.
\ee
for some parameters $A$, $B$, $C$, and $D$ to be determined self-consistently.

For $g>0$, the form of the interaction suggests $O_{m,n}=O_{m+1,n}$, which leads to $C=D=0$.
The interaction Hamiltonian in the mean-field approximation can then be written as
\bea H_{int}= -2g\sum_{m,n}[A +B (-1)^n]i\gamma_{m,n}\gamma_{m,n+1}+g2WL[A^2+B^2]\eea
for a $W\times L$ system.
The Hamiltonian can be diagonalized and the BdG energy level become
\be E_\pm \to \pm \sqrt{(4t\sin k_x)^2+[(4t-8gA )\sin k_y]^2+(8gB \cos k_y)^2}.\label{elD}\ee
with the ground state energy density
\be {E_0\over 2WL}\to g(A^2+B^2)+{1\over (2\pi )^2}\int_0^\pi dk_x\int_{-\pi /2}^{\pi /2}dk_y 
E_-(\vec k),\label{EDel}\ee
Minimizing the energy in the large $g$ limit indicates that both $A$ and $B$ are nonzero, with the symmetry broken states schematically shown in {Fig. \ref{fig7}} for $t>0$.
\begin{figure}[]
\includegraphics[width=3in]{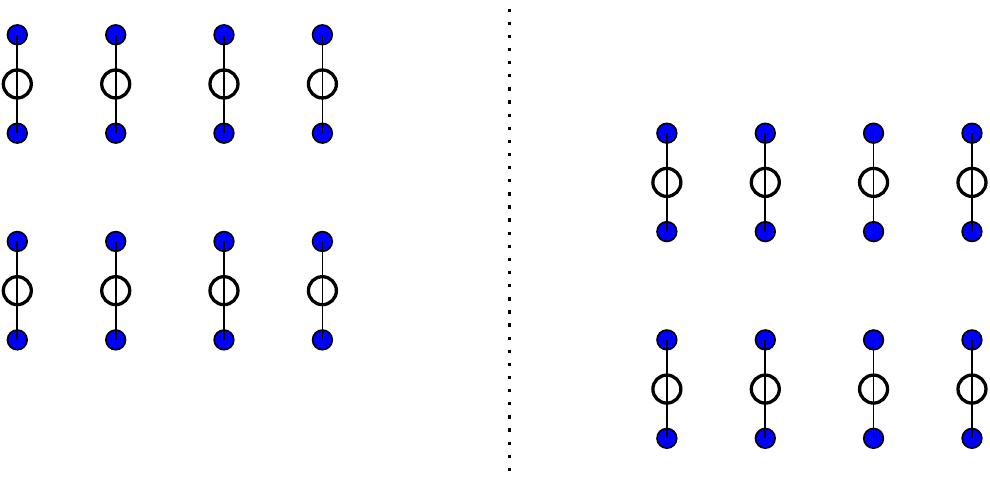}
\caption{Sketch of the two mean-field ground states occurring for $g >> t>0$. The blue dots represent the Majorana modes. The open circles represent unoccupied Dirac fermions from paring the two Majorana modes at the two endpoints of the bond. Similar to the 1D case, the sign of $t$ picks the unoccupied configuration. In addition, there are two equivalent states with Dirac fermions occurring on horizontal bonds and the phase is four-fold degenerate.}
\label{fig7}
\end{figure}

For negative $g$, we expect
\be
\langle i\gamma_{m,n}\gamma_{m,n+1}\rangle=-\langle i\gamma_{m+1,n}\gamma_{m+1,n+1}\rangle
\ee
implying $A=B=0$.
The mean-field interaction Hamiltonian corresponding can be written in a similar way to the $g>0$ case and give the BdG eigenvalues$\pm E_{1,2}$ with
\begin{eqnarray}E_{1,2}^2&=&16(t\sin k_x)^2+(t\sin k_y)^2+(2Dg\cos k_y)^2+(2Cg\sin k_y)^2\\ \nonumber
& \pm& 4t\sqrt{(\sin k_x)^2(D\cos k_y)^2+(\sin^2k_x+\sin^2k_y)(C\sin k_y)^2}.
\end{eqnarray}
The ground-state energy density is then given by ${\cal E}=-g(C^2+D^2)-{1\over (2\pi )^2}\int_0^\pi dk_x\int_0^{\pi /2}dk_y\sum_{i=1}^2E_i(\vec k)$. Once again in the large $|g|$ limit, this mean-field ansatz predicts nonvanishing $C$ and $D$, with eight-fold degenerate states represented in Fig. \ref{fig8}.
\begin{figure}[]
\includegraphics[width=5in]{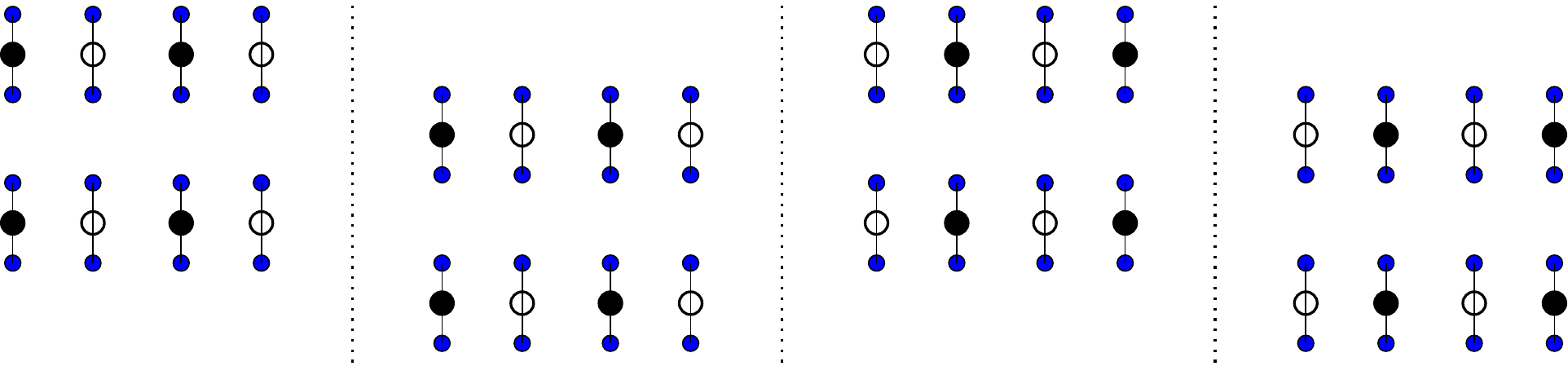}
\caption{Sketch of the mean-field ground states for $-g>>t>0$. The blue dots
represent Majorana fermions sites. The large circles on the bonds indicate filled or empty Dirac fermions formed from the two Majoranas at the endpoints of the bonds. There are four equivalent related by a $\pi/2$ rotation, yielding eight-fold degeneracy.}
\label{fig7}
\end{figure}

\subsection{Weak coupling}
The weak coupling studies shed light on the nature of the quantum phase transitions in this model~\cite{Affleck2017}. Particularly, they suggest that the model for $g>0$ may realize a distinct supersymmetric critical point from the TCI model of the one-dimensional Majorana-Hubbard chain.  We start by labeling the Majorana modes on even and odd rows by $\gamma^{e/o}$. By finding the points with vanishing dispersion, the low-energy modes are given by
 \be \gamma^{e/o}_{\vec r}\approx 2\sqrt{2}[\chi^{e/o+}(\vec r)+(-1)^x\chi^{e/o-}(\vec r)],\label{chid}\ee
 where $\chi^{e/o\pm}(\vec r)$ vary slowly. Combining $\chi^{e\pm}$ and $\chi^{o\pm}$ into $\vec \chi^{+}\equiv (\chi^{e+},\chi^{o+})^T$, $\vec \chi^-\equiv (\chi^{o-},\chi^{e-})^T$, we can write the low-energy noninteracting Hamiltonian as
 \be {\cal H}_0=4it\sum_{\pm}\vec \chi^{T\pm} \cdot [\sigma^z\partial_x+\sigma^x\partial_y]\vec \chi^{\pm}.\ee

 The interaction term in terms of the low-energy fields can be written as
 \be {\cal H}_{int}=64g\chi^{e-}\chi^{e+}\chi^{o+}\chi^{0-}\propto (\bar \chi^+\chi^+)(\bar \chi^-\chi^-),\label{Hinc}\ee
 and is irrelevant in RG.
We note that he theory exhibits an emergent $U(1)$ symmetry for a Dirac fermion obtained by combining the two Majorana modes as
 \be \psi \equiv \left(\begin{array}{c} \chi^{e+}+i\chi^{o-}\\ \chi^{o+}+i\chi^{e-}\end{array}\right).\label{Dirac}\ee
 
The interaction term in the field theory approximation can be written in two ways:
\be {\cal H}_{int}=-64g\psi^\dagger_1\psi^\dagger_2\psi_2\psi_1=32g(\bar \psi \psi )^2.\ee
For $g>0$ this can be exactly rewritten in terms of a complex scalar field $\phi$ and for 
$g<0$ in terms of a real scalar field $\sigma$:
\bea {\cal H}_{int}&\to & m^2|\phi |^2+8m\sqrt{g}(\psi^\dagger_1\psi^\dagger_2\phi +{\rm H.c.})\ \  (g>0)\nonumber \\
&\to& {m^2\over 2}\sigma^2+8m\sqrt{-g}\bar \psi \psi \sigma \ \  (g<0),\eea
through a Hubbard-Stratonovich transformation. Promoting the fields $\phi$ and $\sigma$ to relativistic massive fields, with real-time Lagrangians:
\bea {\cal L}&=&\bar \psi i\gamma^\mu\partial_\mu \psi +|\partial_\mu \phi |^2-m^2|\phi |^2
+g_1(\psi^\dagger_1\psi^\dagger_2\phi +h.c.)\nonumber \\
&-&g_2|\phi |^4,\ \  (g>0)\nonumber \\
&=&\bar \psi i\gamma^\mu\partial_\mu \psi+{1\over 2}(\partial_\mu \sigma )^2-{m^2\over 2}\sigma^2+
g_1\bar \psi \psi \sigma -2g_2\sigma^4.\nonumber \\
&&\ \  (g<0),\label{LFB}\eea
where $64g={g_1^2\over m^2}$ ($-{g_1^2\over m^2}$) for $g>0$ ($g<0$).
  
We expect the transitions in this fermion-boson model (driven by a change in the sign of $m^2$)
to be in the same universality classes as the transitions in the Majorana-Hubbard lattice model. The first 
Lagrangian in Eq. (\ref{LFB}) is the one studied in Refs. \cite{Thomas2005,Lee2007,Zerf2016,Klebanov2016}. It is expected to have a transition into a superfluid 
phase. This transition is believed to be supersymmetric. 
The second Lagrangian in Eq. (\ref{LFB}) is known as  the Gross-Neveu-Yukawa model and discussed in Ref. \cite{Klebanov2016}.
These field theory transitions were shown to be in agreement with mean-field transitions predicted in Ref. \cite{Affleck2017}.

\section{Majorana Hubbard model on the honeycomb lattice}

\begin{figure}[]
\includegraphics[width=3.0in]{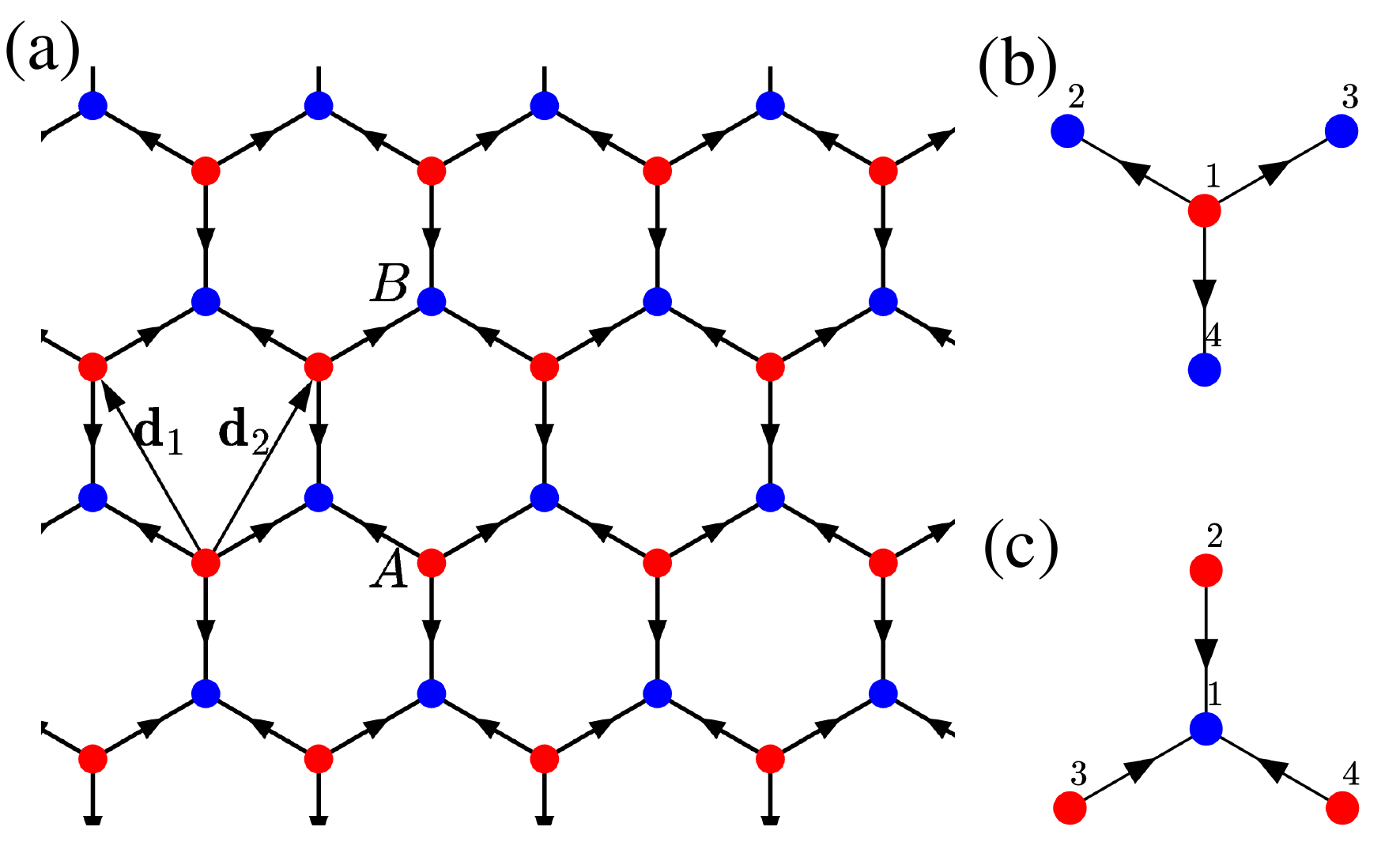}
\caption{(a) Lattice structure of the honeycomb Majorana Hubbard model. The arrows indicate the sign of the hopping terms. (b,c) The ordering of the four operators in $g_1$ and $g_2$ interaction  terms, respectively. (Figure adapted from Ref.\ \cite{Chengshu2018})}
\label{fig9}
\end{figure}
Majorana Hubbard model can be formulated for the honeycomb lattice familiar from the physics of graphene. In this case the most local interaction term involves a site in sublattice A and three adjacent sites in sublattice B (or vice versa) as indicated in Fig.\
\ref{fig9}. The Hamiltonian is given by  $\cH=\cH_0+\cH_{\mathrm{int}}$ with 
\begin{equation}\label{h0}
\cH_0=it\sum_{\langle ij\rangle}\eta_{ij}\alpha_i\beta_j,\ \ \
\cH_{\mathrm{int}}=g_1\sum_{\Ydown}\alpha_i\beta_j\beta_k\beta_l
+g_2\sum_{\Yup}\beta_i\alpha_j\alpha_k\alpha_l.
\end{equation}
Here $\alpha_i$ $(\beta_i)$ denote Majorana operators on the $A$ $(B)$ sublattice,  the phase factors
$\eta_{ij}=\pm 1$ are constrained by the Grosfeld-Stern rule
\cite{Stern2006} and are chosen as indicated in Fig.\,\ref{fig9}(a).

In the absence of interactions ($g_1=g_2=0$) the model exhibits a unique
ground state with linearly dispersing single particle
excitations near the $\pm\bK$ corners of the
hexagonal Brillouin zone, analogous to graphene
\cite{Neto_RMP}. The gapless nature of the excitations in $\cH_0$ is protected by the combination of inversion $\cP$ and time reversal symmetry $\cT$. The latter is generated by $(\alpha_j,\beta_j)\to (\alpha_j,-\beta_j)$ and $i\to -i$. It is easy to see that interaction terms in Eq.\ (\ref{h0}) break $\cT$ because they contain an odd number of $\beta$ operators. One might therefore expect that, unlike in the square lattice model, even weak interactions could significantly change the nature of the ground state and the low-energy excitations. Indeed, analysis performed in Ref.\ \cite{Chengshu2018} shows that arbitrarily weak interactions open up a gap in the spectrum of the model.

\begin{figure}
\includegraphics[width=9.7cm]{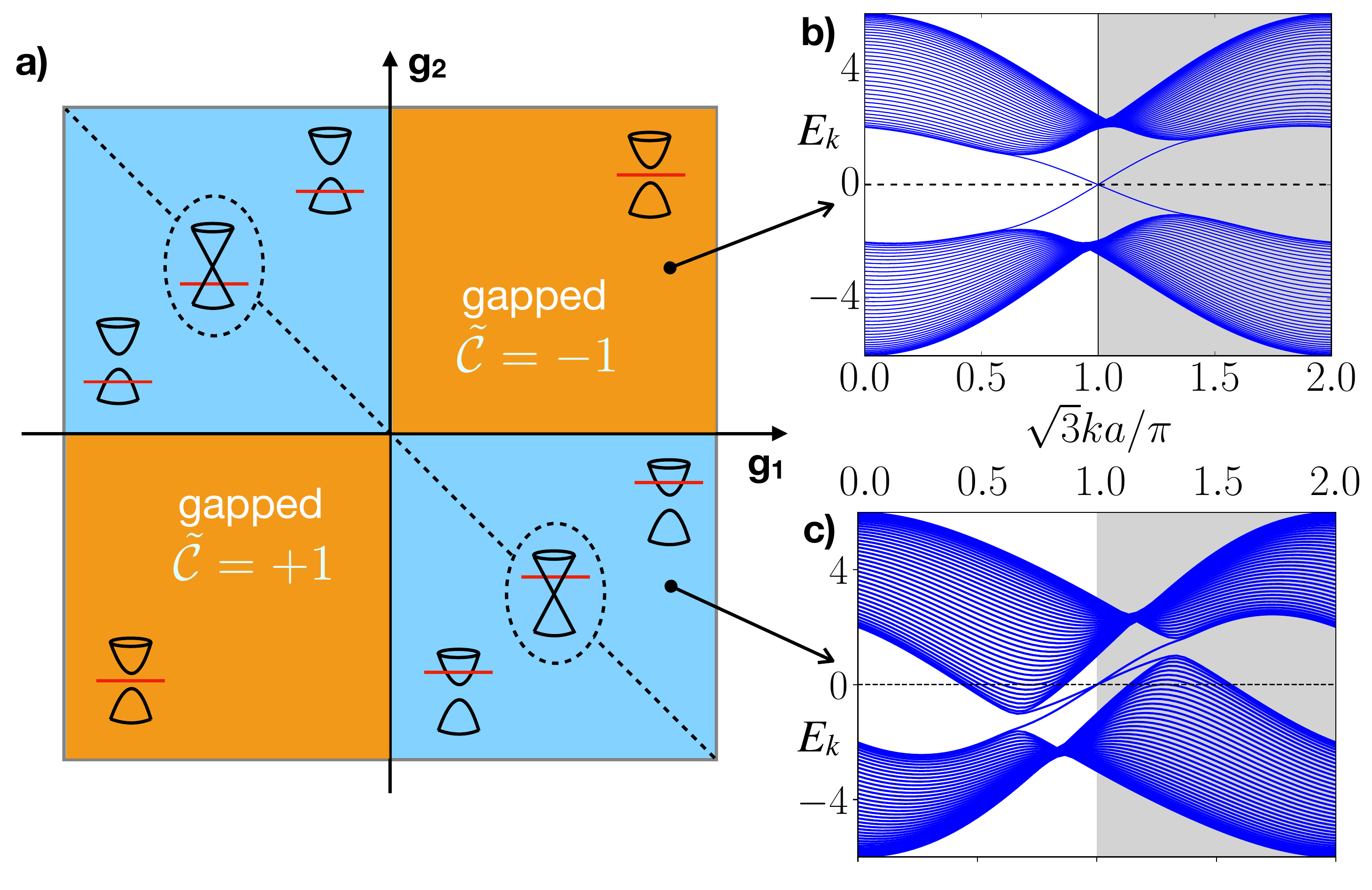}
\caption{(a) The phase diagram of the Majorana Hubbard model on the honeycomb lattice at
  weak to intermediate coupling. MF Hamiltonian Eq.\ (\ref{hmf}) on a strip 
with the zig-zag boundary (with $\tau_1=\pm0.15,\ \tau_2=0.1$) is
used to illustrate  the energy spectra 
 in the gapped Majorana Chern insulator phase (b) and the gapless
  Majorana metal phase (c). (Figure adapted from Ref.\ \cite{Chengshu2018})}\label{fig10}
\end{figure}

To see this it is instructive to write down the low-energy effective theory by expanding the Majorana
fields around the two nodal points at $\pm\bK$ (see Ref.\ \cite{Chengshu2018}  for details).  One thus obtains
\begin{equation}\label{hc0}
\cH_0\simeq -v\int \rd^2r\sum_{\sigma=\pm}\sigma\left(\alpha_{
    \sigma}\partial_\sigma\beta_{\bar\sigma}+ \beta_{\bar\sigma}\partial_\sigma\alpha_{
    \sigma}\right),
\end{equation}
where $(\alpha_\pm,\beta_\pm)$
are the long-wavelength components of the Majorana fields near points
$\pm\bK$, ${\bar \sigma}=-\sigma$, $v=3ta$ is the
characteristic velocity, $a$ denotes the lattice constant, and
$\partial_\pm=(\partial_x\pm i\partial_y)$.  Similarly one finds
\begin{equation}\label{hc1}
\cH_{\rm int}\simeq {24\sqrt{3}}a\int
d^2r\sum_{\sigma=\pm}\bigl[g_1\beta_{
    \sigma}\beta_{\bar\sigma}(\alpha_{
                                \sigma}\partial_{\sigma}\beta_{\bar\sigma}) 
+ g_2 \alpha_{\bar\sigma} \alpha_{\sigma} (\beta_{\bar\sigma}\partial_\sigma\alpha_{
    \sigma})\bigr]. \nonumber
\end{equation}
Power counting indicates that $\cH_{\rm int}$ is even more {\em irrelevant} than the interaction term on the square lattice due to the extra derivative. Nevertheless it is not innocuous. The key point to note is that terms in parentheses in Eq.\ (\ref{hc1}) coincide with the terms present in $\cH_0$. Therefore these terms will have non-zero expectation value in the ground state of the non-interacting system. As weak interactions are turned on it is permissible to replace these terms in  $\cH_{\rm int}$ by their expectation value.  The low-energy  Hamiltonian then becomes 
\begin{equation}\label{hc2}
\cH\approx-\int \rd ^2r\sum_{\sigma=\pm} \sigma\Psi^\dagger_\sigma
\begin{pmatrix}
-g_2M & -v\partial_{\bar\sigma} \\
v\partial_{\sigma} & g_1M 
\end{pmatrix}
\Psi_\sigma,
\end{equation}
where $\Psi_\sigma=(\alpha_\sigma,\beta_\sigma)^T$, $M=24\sqrt{3}am$ and $m=\sigma\langle\alpha_{
  \sigma}\partial_\sigma\beta_{\bar\sigma}\rangle
=\sigma\langle\beta_{\bar\sigma}\partial_\sigma\alpha_{ \sigma}\rangle$.
 Assuming translation invariance the spectrum of $\cH$
is easily obtained by passing to the momentum representation,
\begin{equation}\label{hc3}
E_{k,\sigma}={1\over 2}\sigma M(g_1-g_2)\pm\sigma\sqrt{v^2k^2+{1\over 4}M^2(g_1+g_2)^2},
\end{equation}
where $k=(k_x^2+k_y^2)^{1/2}$.  Arbitrarily weak interactions open a gap (except when $g_1=-g_2$). In addition unequal interaction strengths lead to an offset in energies between the two nodal points at $\pm\bK$. The resulting phase diagram of the model at weak coupling is given in Fig.\ \ref{fig10}(a).  

The weakly interacting model thus supports gapped and gapless phases. The gapped phases occur for $g_1g_2>0$, are topologically nontrivial and are characterized by Chern number $\tilde{\cC}=-{\rm sgn}(g_1)$. In the geometry with an open boundary they support chiral Majorana edge modes as illustrated in Fig.\ \ref{fig10}(b). Gapless phases occur for $g_1g_2<0$ and support antichiral edge modes illustrated in Fig.\ \ref{fig10}(c). These  propagate in the same
direction on two opposite edges and are protected by their real-space
segregation from the bulk modes \cite{Colomes2018}. 

At weak to moderate coupling mean-field (MF) theory provides a good description of the interacting Hamiltonian. On the lattice we may approximate $\alpha_i\beta_j\beta_k\beta_l\to
\langle\alpha_i\beta_j\rangle\beta_k\beta_l$ where the expectation
value lives on the nearest neighbor bond and coincides with terms already present in
$\cH_0$. The operator product  $\beta_k\beta_l$ then describes coupling between next
nearest neighbors.  This motivates considering the MF Hamiltonian with first and
second neighbor hoppings
\begin{equation} \label{hmf}
\cH_{\MF}=i\sum_{\langle ij\rangle}\eta_{ij}\tau_0\alpha_{i}\beta_{j}+i\sum_{\langle\langle ij\rangle\rangle}\eta_{ij} (\tau_1\alpha_{i}\alpha_{j}+\tau_2\beta_{i}\beta_{j}),
\end{equation}
with the signs specified in Fig.\,\ref{fig9}(a). MF theory uses the 
eigenstates of  $\cH_{\MF} $ as variational
wavefunctions for the full Hamiltonian (\ref{h0}), parametrized by
$\{\tau_a\}$.  Minimizing the ground state energy $\langle\Psi_{\MF}|\cH|\Psi_{\MF}\rangle$  with respect to these variational parameters leads to self-consistent MF gap equations 
\begin{equation}\label{sc}
\begin{split}
\tau_0&=t+g_1\Delta_2+g_2\Delta_1,\\
\tau_1&=g_2\Delta_0,\\
\tau_2&=g_1\Delta_0,
\end{split}
\end{equation}
where the order parameters $\Delta_0=i\langle\alpha_1\beta_2\rangle$,
$\Delta_1=i\langle\alpha_1\alpha_2\rangle$ and
$\Delta_2=i\langle\beta_1\beta_2\rangle$ are defined on first and second neighbor bonds of the honeycomb lattice. 

Numerical solution of the gap equations (\ref{sc}) confirms the structure of the phase diagram given in Fig.\ \ref{fig10}(a) for weak to intermediate coupling. In addition, unbiased exact diagonalization (ED) study of the full interacting Hamiltonian $\cH$ shows an excellent agreement with the MF theory for $|g_{1,2}|\lesssim 2 t$; specifically order parameters $\Delta_a$ calculated from ED are found to be in  qualitative and quantitative agreement with the solution of the MF equations (\ref{sc}). At stronger coupling ED and MF results begin to diverge signaling the breakdown of the MF theory in this limit. The nature of the ground state at strong coupling is not known at present. For $t=0$ the Hamiltonian can be mapped via a Jordan-Wigner transformation onto a spin-${1\over 2}$ model on a triangular lattice with three-spin interaction terms on each triangle \cite{Chengshu2018}.   The spin Hamiltonian is frustrated and ED suggests a doubly degenerate ground state with a gap to excitations which grows linearly with $g$ for $g>5t$. The strong coupling limit of the Majorana Hubbard model on the honeycomb lattice thus presents an interesting open problem.

\section{Majorana surface code through interacting Majoranas on the honeycomb lattice}

An mentioned above, direct braiding of Majorana zero modes is not enough for universal quantum computing. The Fibonacci anyons of the previous section provide one solution. The ``surface codes" \cite{Surf_Code_Kitaev, Surf_Code_Freedman} provide an alternative approach to universal quantum computation. In surface codes, qubits are arranged in a two-dimensional array on a surface. 
Kitaev's celebrated toric code~\cite{kitaev}, which allows for fault-tolerant quantum computing, is the basis from which the theory of surface codes has evolved. While the toric code relied on the nontrivial topology of the torus provided by the periodic boundary conditions, other versions known as planar surface codes (or simply planar codes) were developed \cite{Surf_Code_Kitaev, Surf_Code_Freedman}. There are certain versions of the toric code, which can realize all the required gates for universal topological quantum computing~\cite{Dennis2002,Fowler2012}.

In this scheme, states are transformed by the projective measurement of operators known as stabilizers \cite{Raussendorf, Raussendorf_2, Fowler_Martinis, Fowler_Surface_Code}. The logical qubits are encoded in certain anyon charges and tailored measurement sequences of stabilizers effectively braid the logical qubits. Remarkable experimental progress has been made on surface codes with superconducting qubits \cite{Martinis, Martinis_Surf_Code_Array, Steffen_Surf_Code_Array}.

Vijay, Hsieh, and Fu showed that an exactly solvable model of interacting Majorana fermions with six-Majorana interactions \cite{bravyi,VHF} realizes the $Z_{2}$ topological order and has three fundamental anyonic excitations, which allow for a Majorana surface code. They also proposed a concrete physical realization of the model based on Josephson-coupled mesoscopic topological superconductors, and after identifying the logical qubits demonstrated the implementation of the CNOT, T, and the Hadamard gates in the proposed surface code, showing that it allows for universal quantum computing. Here we present a basic introduction to this model of interacting Majoranas and its anyonic excitations. For a detailed discussion of the physical implementation and the logical gates, we refer the reader to Ref. \cite{VHF}.

Majorana fermions form a honeycomb lattice and the Hamiltonian is 
\begin{align}\label{eq:Hamiltonian}
H = -u\sum_{p}\mathcal{O}_{p} \hspace{.25in} \mathcal{O}_{p} \equiv i\prod_{n \in\rm{vertex} (p)}\gamma_{n},
\end{align}
where $p$ represents a hexagonal plaquette and $\mathcal{O}_{p}$ is the product of six Majorana operators around the plaquette. Two different plaquettes can have either two or no common Majoranas. Thus $[\mathcal{O}_{p},\mathcal{O}_{p'}]=0$. As $\mathcal{O}_{p}$ squares to the identity, it can have $\pm1$ eigenvalues and for $u>0$, the ground states satisfy
\be
\mathcal{O}_{p}|{\Psi_{0}}\rangle = |{\Psi_{0}}\rangle,  \label{p}
\ee
for all $p$.

\begin{figure}[h]
\includegraphics[width=3in]{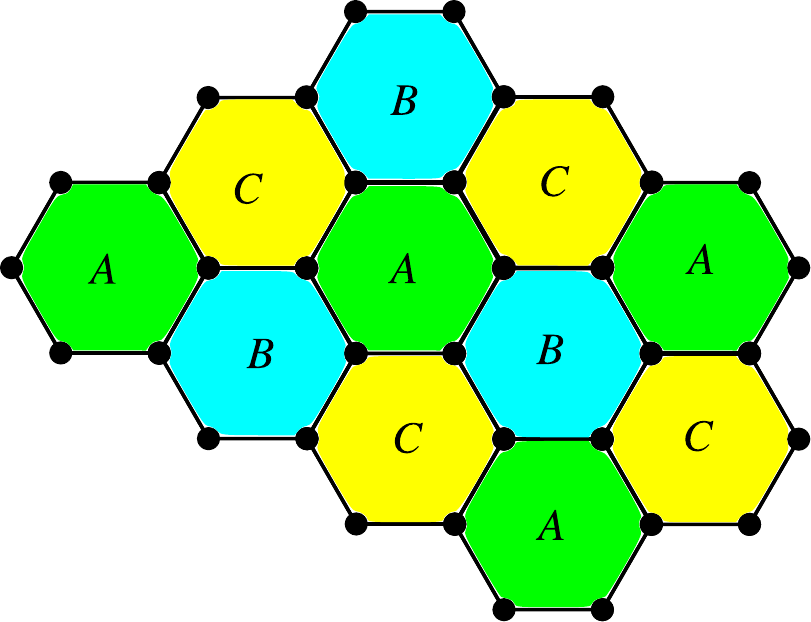}
\caption{The three types of fundamental plaquettes in the Vijay-Hsieh-Fu model.}
\label{fig:plaq}
\end{figure}

A counting argument shows that the ground state must be four-fold degenerate on the torus. First the total Fermion parity $\Gamma=(i)^{N/2}\prod_{n=1}^N\gamma_n$. Fixing the fermion parity gives a $2^{N/2-1}$-dimensional Hilbert space. As shown in Fig.\ \ref{fig:plaq}, the plaquettes on the honeycomb lattice can be divided into three types A, B, and C, such that the plaquettes of each type include all the sites on the lattice and no site belongs to more than one plaquette of one type. Thus 
\begin{align}\label{eq:type}
\Gamma = \prod_{p\in A}\mathcal{O}_{p} = \prod_{p\in B}\mathcal{O}_{p} = \prod_{p\in C}\mathcal{O}_{p}
\end{align}
The operators $\mathcal{O}_{p}$ on each plaquette type fix one third of the plaquette eigenvalue conditions of the ground state. Taking into account the fermion parity, the plaquettes of each type give $2^{N/6 - 1}$ constraints. Therefore the degeneracy for ground state is expected to be $2^{N/2-1}/(2^{N/6 - 1})^3=4$, the same as a $Z_2$ topological phase. The four ground states can only be distinguished by nonlocal Wilson loop operators.

Excitations above the ground sate manifold are gapped and correspond to flipping the sign of the $\mathcal{O}_{p}$ eigenvalue on plaquettes. Fermion parity conservation requires that the plaquette eigenvalues are flipped in pairs, which can be achieved by string operators that flip the fermion parity on the edges of plaquettes. Due to Eq. \eqref{eq:type}, these pairs of excited plaquettes must belong to the same ABC type. It is not possible to change one excitation type to another, and the model supports \textit{three} types of elementary excitations, labeled by $A$, $B$ and $C$ plaquette type. 

Vijay, Hsieh, and Fu determined the statistics of these three type of excitations by considering various braiding processes, implemented by acting with bond parity operators $i\gamma_{n}\gamma_{m}$ to move the excitations around. All three types of plaquette excitations have boson self-statistics and mutual semion statistics. They also studied the statistics of composite excitations $AB$, $BC$, $AC$ and $ABC$ with results summarized in the table below:
\begin{align*}
  \begin{tabular}{|c|c|c|c|c|c|c|c|c|}
  \hline
    & $1$ & $A$ & $B$ & $C$ & $AB$ & $BC$ & $AC$ & $ABC$\\
    \hline
$1$ & $+1$ & $+1$ & $+1$ & $+1$ & $+1$ & $+1$ & $+1$ & $+1$\\ 
    \hline
$A$ & $+1$ & $+1$ & $-1$ & $-1$ & $-1$ & $+1$ & $-1$ & $+1$\\
    \hline  
$B$ & $+1$ & $-1$ & $+1$ & $-1$ & $-1$ & $-1$ & $+1$ & $+1$\\
    \hline
$C$ & $+1$ & $-1$ & $-1$ & $+1$ & $+1$ & $-1$ & $-1$ & $+1$\\
    \hline
$AB$ & $+1$ & $-1$ & $-1$ & $+1$ & $-1$ & $-1$ & $-1$ & $+1$\\
    \hline
$BC$ & $+1$ & $+1$ & $-1$ & $-1$ & $-1$ & $-1$ & $-1$ & $+1$\\
    \hline
$AC$ & $+1$ & $-1$ & $+1$ & $-1$ & $-1$ & $-1$ & $-1$ & $+1$\\
    \hline
$ABC$ & $+1$ & $+1$ & $+1$ & $+1$ & $+1$ & $+1$ & $+1$ & $-1$\\
    \hline
  \end{tabular}
\end{align*}

The commuting operators $\mathcal{O}_{p}$ serve as the stabilizers of the Majorana surface code. A physical realization relies on placing an array of Josephson-coupled superconducting islands on the surface of a topological insulator. Surface codes rely on the projective measurement of these stabilizers, which can be achieved by decreasing the Josephson coupling of an island to activate quantum phase slips. The charging energy of the island then creates an energy difference between different stabilizer eigenstates.   Furthermore, the energy gaps between certain levels of the island are sensitive to the stabilizer eigenvalue. Thus measuring the energy gap, e.g., by shining a microwave beam  and measuring the phase shift of the transmitted photons, effectively performs a measurement of the stabilizer operator \cite{Top_Transmon, Schoelkopf}.

\section{Sachdev-Ye-Kitaev model and its realizations}

A class of models with {\em random} all-to-all interactions between $N$ fermions, originally introduced by Sachdev and Ye in 1996 \cite{SY1996}, came to prominence recently when Kitaev noticed their intriguing connections to the physics of black holes and quantum chaos \cite{Kitaev2015,Maldacena2016}. The resulting Sachdev-Ye-Kitaev (SYK) model is defined by the Hamiltonian 
\begin{equation}\label{hsyk}
H_{\rm SYK}=\sum_{i<j<k<l} J_{ijkl}\gamma_i\gamma_j \gamma_k\gamma_l,
\end{equation}
where $J_{ijkl}$ are real random independent coupling constants drawn from a Gaussian ensemble with 
\begin{equation}\label{syk1}
\overline{J_{ijkl}}=0, \ \ \ \overline{J_{ijkl}^2}={3!\over N^3}J^2,
\end{equation}
and $J$ is the characteristic energy scale. We observe that $H_{\rm SYK}$ has the general structure of an interacting Majorana model Eq.\ (\ref{eq:H_m}) with bilinear terms absent and interactions connecting all fermions. The model can thus be considered zero-dimensional -- there is no sense of distance between fermions.

Due to its relevance to fundamental unsolved questions in string theory, quantum gravity, information theory and quantum chaos there  now exists a voluminous literature on the SYK model proper \cite{Sachdev2015,Xu2016,Polchinski2016,Verbaar2016,Kitaev:2017awl}, its various extensions \cite{Gu2016,Fu2016,Berkooz2016,Hosur2016,Witten2016,Altman2016,Bi2017,Lantagne2018} and proposed physical realizations \cite{Danshita2017,Danshita2017b,Garcia2017,Laflamme2017,Pikulin2017,Alicea2017,achen2018}. The aim of this Section is to review some of its basic physical properties, outline its remarkable connections to other areas of physics and briefly discuss proposed experimental realizations.  A more comprehensive treatment of this subject can be found in two recent review articles \cite{Rosenhaus_rev,Rozali_rev}.

\subsection{Basic properties of the model}

Perhaps the most remarkable property of the SYK model is that, despite being maximally strongly interacting, it is in fact exactly solvable in the limit of large $N$.  The solution (see e.g.\ Refs. \cite{Kitaev2015,Maldacena2016} for details) proceeds by writing down the imaginary-time action $S_{\rm SYK}$ associated with Hamiltonian (\ref{eq:H_m}) and then averaging over random couplings $J_{ijkl}$ using the replica formalism. At large $N$ the resulting effective action $S_{\rm eff}$ is seen to be dominated by the saddle point contribution. This is to say that, remarkably, mean-field theory provides the exact solution at large $N$. The resulting mean-field theory is however highly nontrivial in that it describes a non-Fermi liquid ground state with a number of interesting properties.

The relevant saddle point equations (obtained by minimizing $S_{\rm eff}$) relate the averaged fermion propagator 
\begin{equation}\label{syk2}
G(\tau,\tau')={1\over N}\sum_j\langle {\cal T}_\tau\gamma_j(\tau)\gamma_j(\tau')\rangle
\end{equation}
to the self energy $\Sigma(\tau)$ as follows
\begin{equation}\label{syk3}
G(\omega_n)=[-i\omega_n-\Sigma(\omega_n)]^{-1}, \ \ \ \ \ \Sigma(\tau)=J^2G^3(\tau).
\end{equation}
Here $G(\omega_n)=\int_0^\beta d\tau e^{i\omega_n\tau}G(\tau)$ and
$\beta=1/k_BT$ is the inverse temperature. At non-zero temperatures
the propagator and the self energy are defined for discrete Matsubara
frequencies $\omega_n=\pi T(2n+1)$ with $n$ integer and taking
$k_B=1$ here and henceforth. We also assume time-translation invariance, $G(\tau,\tau')=G(\tau-\tau')$. 

The innocent looking pair of equations (\ref{syk3}) engenders considerable structure. At low frequencies, when we can neglect $i\omega_n$ compared to $\Sigma(\omega_n)$, they are invariant under time reparametrization of the form 
\begin{equation}\label{syk4}
G(\tau,\tau')\to [f'(\tau) f'(\tau')]^{1/4}G(f(\tau),f(\tau')), \ \ \ \Sigma(\tau,\tau')\to  [f'(\tau) f'(\tau')]^{3/4}\Sigma(f(\tau),f(\tau')),
\end{equation}
for an arbitrary smooth function $f(\tau)$. This reparametrization invariance already hints at deep connections with string theory and black hole physics. More practically, Eqs.\ (\ref{syk4}) can be used to extract the asymptotic behavior of the propagator in the low-frequency ``conformal'' regime, 
\begin{equation}\label{syk5}
G_c(\omega_n)=i\pi^{1/4}{{\rm sgn}(\omega_n)\over\sqrt{J|\omega_n|}}
\end{equation}
and by analytic continuation also the corresponding zero-temperature spectral function 
\begin{equation}\label{syk6}
A_c(\omega)={1\over \sqrt{2}\pi^{3/4}}{1\over \sqrt{J|\omega|}}.
\end{equation}
These expressions are valid for $|\omega|\ll J$ and must cross over to
the $1/\omega$ behavior at large frequencies. The characteristic square-root singularity present in the spectral function clearly indicates the non-Fermi liquid nature of the ground state and the absence of quasiparticle excitations.

Another interesting property of the model that readily follows from these considerations is the non-vanishing ground state entropy per particle 
\begin{equation}\label{syk7}
{S_0\over N}=\lim_{T\to 0}{S(T)\over N}\simeq {1\over 2}\ln{2}-{\pi^2\over 64}\simeq 0.19236. 
\end{equation}
Although the ground state itself is non-degenerate  the extensive entropy arises from the exponentially large number of excited states that exist close to the ground state.

The SYK model also exhibits intriguing connections to quantum chaos. The central concept quantifying the amount of chaos in a quantum system is {\em scrambling}. This refers to the process in which quantum information deposited in the system locally gets distributed among all its degrees of freedom. Black holes are thought to scramble with the maximum possible efficiency: they saturate the fundamental bound on the relevant Lyapunov exponent defined below. To describe scrambling quantitatively it is customary to define an out-of-time-order correlator (OTOC), as 
\begin{equation}\label{syk8}
F_{ij}(t)=\langle\gamma_j(t)\gamma_i(0)\gamma_j(t)\gamma_i(0)\rangle.
\end{equation}
For black holes in
Einstein gravity scrambling occurs exponentially fast with $1-F(t)\sim
e^{\lambda_L t}/N$ where the growth rate is given by the maximal Lyapunov
exponent $\lambda_L=2\pi T$ \cite{Maldacena2016b}. Similarly, for the SYK model in the
large-$N$ limit one finds \cite{Kitaev2015,Maldacena2016}
\begin{equation}\label{syk11}
1-F(t)\sim {J\over NT}e^{\lambda_Lt}.
\end{equation}

The emergent conformal invariance at low energies, extensive ground state entropy, and the behavior of OTOC all hint at a subtle relationship between the SYK model and a black hole. The large-$N$ analysis is indeed thought to  establish the SYK Hamiltonian (\ref{hsyk}) as a paradigmatic example of a broad class of systems called  holographic quantum matter. Such systems generally lack conventional quasiparticle description and are studied by means of holographic dualities which connect strongly coupled quantum theories with their gravitational duals.

\subsection{Extensions and variants of SYK}

Of many proposed extensions and variants of the SYK model we only have space to mention few. The complex-fermion version, also known as cSYK or Sachdev-Ye (SY) model, is defined by 
\begin{equation}\label{syk20}
H_{\rm cSYK}=\sum_{ij;kl} J_{ij;kl}c^\dagger_ic^\dagger_jc_kc_l -\mu\sum_jc^\dagger_jc_j,
\end{equation}
where $c^\dagger_j$ creates a spinless fermion and $ J_{ij;kl}$ are
zero-mean complex random variables satisfying $ J_{ij;kl}=J^*_{kl;ij}$ (hermiticity)
and $J_{ij;kl}=-J_{ji;kl}=-J_{ij;lk}$ (antisymmetry).  At half filling $(\mu=0)$ the cSYK model exhibits essentially the same physical properties as the SYK model discussed above. Away from half filling $(\mu\neq 0)$ it is known to develop an interesting spectral asymmetry  and correlators that match with those of charged extremal black holes \cite{Sachdev2015}.

Novel behaviors have been noted in models described by the SYK Hamiltonian Eq.\ (\ref{hsyk}) with modified coupling constants. Taking 
\begin{equation}\label{syk21}
J_{ijkl}=-{1\over 8}\sum_{a=1}^MC_{a[ij}C_{kl]a}
\end{equation}
where $C_{aij}$ are independent random variables and angular brackets denote antisymmetrization. For $M=N$ one obtains a supersymmetric SYK model \cite{Fu2016} with a single supercharge $Q=i\sum_{i<j<k}C_{ijk}\gamma_i\gamma_j\gamma_k$. Changing the upper bound in the summation (\ref{syk21}) to $M=pN$ generates a family of SYK-like non-Fermi liquids with a  continuously varying fermion scaling dimension $\Delta_f$ as a function of parameter $p$ \cite{Bi2017}. A phase transition into a Fermi liquid phase can be induced in the cSYK model Eq.\ (\ref{syk20}) by coupling it to a system of `peripheral' fermions $\psi_a$ whose Hamiltonian only contains bilinear terms \cite{Altman2016}.

Extensions of the SYK and cSYK model to higher spatial dimensions $d$ also show interesting physics. Most approaches consider a $d$-dimensional spatial lattice of islands, each described by $H_{\rm cSYK}$ or $H_{\rm SYK}$,  coupled to one another in various ways. When the coupling between neighboring islands is via random 4-fermion terms the resulting higher-dimensional system retains many attributes of the SYK model and shows new properties unique to $d>0$ such as diffusive energy transport and emergent ``butterfly velocity'' describing the propagation of chaos in space \cite{Berkooz2016,Gu2016,Gu2017,Jian2017}.  Tunneling between islands, described by random 2-fermion terms, leads to a strongly correlated metal with intriguing properties such as linear in temperature resistivity, reminiscent of the high-$T_c$ cuprates and other strongly correlated  systems \cite{Balents2017}. Other types of inter-island coupling schemes have been considered leading to interesting extensions including non-Fermi liquid metal phases \cite{Greevy2017} and topological states \cite{Zhang2018}.

Also worth mentioning are randomness-free variants of the SYK model \cite{Witten2016,Klebanov2017} which have been employed recently  as a basis to model the strange metal phase in high-$T_c$ cuprates \cite{Wu2018}.

\subsection{Experimental realizations}

Given its remarkable properties it would be of obvious interest to develop an experimental realization of the SYK Hamiltonian (\ref{hsyk}) or one of its variants. Several realizations have been proposed in the recent literature. Early on a realization of the cSYK model 
has been proposed using ultracold gases
\cite{Danshita2017}.  A protocol for digital quantum simulation of
both the complex and Majorana fermion versions of the model has been
discussed \cite{Garcia2017}.  A first step towards a quantum simulation has been implemented on a four-qubit simulator exploiting molecules of trans-crotonic acid \cite{Laflamme2017}. 

Three different solid-state realizations have been discussed. Two of these are based on Majorana fermions in the Fu-Kane superconductor and in semiconductor quantum wires \cite{Pikulin2017,Alicea2017}. One employs ordinary complex fermions in the lowest Landau level (LL$_0$) of a graphene flake with an irregular boundary \cite{achen2018}. Two key challenges must be overcome by any solid-state implementation: (i) a way must be found to suppress bilinear terms (of the form $it_{ij}\gamma_i\gamma_j$) which are normally dominant in any natural system of $N$ Majorana or complex fermions, and (ii) at the same time significant four-fermion interactions that are sufficiently random and all-to-all must be present. The proposed realizations \cite{Pikulin2017,Alicea2017,achen2018} rely on approximate symmetries of the solid-state system that disallow fermion bilinears but permit interaction terms. For instance the graphene flake proposal \cite{achen2018} takes advantage of the chiral symmetry of electrons hopping between sites of the honeycomb lattice which guarantees that, at the non-interacting level, LL$_0$ remains perfectly degenerate in the presence of strong symmetry preserving disorder. Coulomb interaction, when expressed in the single-particle basis of this strongly disordered system, produces random matrix elements that are all-to-all for a small flake, precisely as required to form the cSYK Hamiltonian.

\section{Conclusions}
Interacting Majorana fermion systems open a new and unexpected chapter in the book on strongly correlated quantum matter.  Their novelty derives from the basic algebraic properties of the Majorana operators encapsulated in Eq.\ (\ref {eq:comm}). 
 These imply that in any lattice model the simplest interaction term one can write contains four distinct lattice sites. This is to be contrasted with the standard Hubbard model for spinfull complex fermions (electrons) which contains only on-site interactions. The second important feature concerns model symmetries. While any model Hamiltonian given in terms of Majorana operators can be rewritten in complex fermion basis (and sometimes also in spin-${1\over 2}$ operator basis) such transformations typically give very complicated and unnatural looking Hamiltonians. The basic Majorana-Hubbard chain discussed in Sec.\ III is a good example of this rule. As a result model Hamiltonians that appear very natural in the Majorana basis correspond to complex fermion or spin Hamiltonians that have often not been previously studied and can thus yield new physics.

A good example of this new physics is the emergent supersymmetry realized in the vicinity of a critical point in interacting  1D Majorana models discussed in Sec.\ III and IV. While the same physics is known to occur at the tricritical Ising point in certain spin models one has to vary at least two independent parameters to tune the spin system to achieve criticality in those systems. In Majorana models, owing to their distinct symmetries, only a single parameter needs to be tuned  making the potential experimental realization of supersymmetry much easier. Another good example is the emergent black hole physics in the SYK model reviewed in Sec.\ X. While a spin model can be written with similar properties \cite{SY1996} great effort must be expended to avoid the competing spin glass phase in that setting. By contrast the Majorana version introduced by Kitaev \cite{Kitaev2015} does not suffer this complication and the scale-invariant correlated liquid emerges as the natural ground state.

Theoretical developments described in this review have been inspired by the remarkable experimental progress over the last five years in engineering and probing solid-state systems with unpaired Majorana zero modes.  This progress has in turn been precipitated by the pioneering theoretical works outlining this possibility \cite{Kitaev2001,FuKane2008,Sau2010,Alicea2010,Lutchyn2010,Oreg2010}.  Thus in 2012 Majorana zero modes have been observed for the first time in the seminal Delft experiment on InSb quantum wires \cite{Mourik2012}. Although the quantum wire geometry remains the most developed and studied platform to date \cite{Das2012,Deng2012,Rokhinson2012,Finck2013,Hart2014,Nadj-Perge2014,Deng2016,Zhang2018x}, signatures of Majoranas have been also reported in other physical systems including the Fu-Kane superconductor realized at the interface between a topological insulator Bi$_2$Te$_3$ and superconductor NbSe$_2$ \cite{Jia2016a,Jia2016b}. More recently, Majorana fermions have been reported in surfaces of the iron based superconductor FeTe$_{0.55}$Se$_{0.45}$ \cite{Wang2018}. 

The above developments bring closer to reality the possibility to experimentally observe interaction effects in systems composed of unpaired Majorana fermions. 
Importantly for these efforts the ability to tune the chemical potential in the proximitized topological insulator surface state has been demonstrated \cite{Cho2013}. As emphasized in Sec. II such an ability is required to bring about the regime of strong interactions between Majoranas by suppressing the otherwise dominant tunneling terms. It would therefore appear that all the necessary ingredients to begin systematic experimental exploration of strongly interacting phases of Majorana fermions are falling into place. We hope that this article will help stimulate interest in these explorations.

\section*{Acknowledgements}
The authors' understanding of the subject has been shaped by conversations and correspondence with many colleagues. Of these we would like to acknowledge I. Affleck, J. Alicea, C. Beenakker, P. Fendley, Liang Fu, E. Lantagne-Hurtubise, Chengshu Li, Tianyu Liu, D.I. Pikulin,  and S. Sachdev. The work presented in this article was supported by NSERC and by CIfAR.

\bibliography{maj}

\end{document}